\documentclass[preprint2]{aastex6}
\usepackage{natbib}
\usepackage{graphicx}
\usepackage[caption=false]{subfig} 
\usepackage{booktabs}
\usepackage{hyperref}
\usepackage{textgreek}
\usepackage{soul}
\usepackage{amssymb}

\begin{document}
\title{The JCMT Gould Belt Survey: Evidence for Dust Grain Evolution in Perseus Star-forming Clumps}

\author{Michael Chun-Yuan Chen\altaffilmark{1}}
\author{J. Di Francesco\altaffilmark{2, 1}}
\author{D. Johnstone\altaffilmark{3, 2, 1}}
\author{S. Sadavoy\altaffilmark{4}}
\author{J. Hatchell\altaffilmark{5}}
\author{J.C. Mottram\altaffilmark{6}}
\author{H. Kirk\altaffilmark{2}}
\author{J. Buckle\altaffilmark{7, 8}}
\author{D.S. Berry\altaffilmark{3}}
\author{H. Broekhoven-Fiene\altaffilmark{1}}
\author{M.J. Currie\altaffilmark{3}}
\author{M. Fich\altaffilmark{9}}
\author{T. Jenness\altaffilmark{3, 10}}
\author{D. Nutter\altaffilmark{11}}
\author{K. Pattle\altaffilmark{12}}
\author{J.E. Pineda\altaffilmark{13, 14, 15}}
\author{C. Quinn\altaffilmark{11}}
\author{C. Salji\altaffilmark{7, 8}}
\author{S. Tisi\altaffilmark{9}}
\author{M.R. Hogerheijde\altaffilmark{6}}
\author{D. Ward-Thompson\altaffilmark{12}}
\author{P. Bastien\altaffilmark{16}}
\author{D. Bresnahan\altaffilmark{12}}
\author{H. Butner\altaffilmark{17}}
\author{A. Chrysostomou\altaffilmark{18}}
\author{S. Coude\altaffilmark{16}}
\author{C.J. Davis\altaffilmark{19}}
\author{E. Drabek-Maunder\altaffilmark{20}}
\author{A. Duarte-Cabral\altaffilmark{5}}
\author{J. Fiege\altaffilmark{21}}
\author{P. Friberg\altaffilmark{3}}
\author{R. Friesen\altaffilmark{22}}
\author{G.A. Fuller\altaffilmark{14}}
\author{S. Graves\altaffilmark{3}}
\author{J. Greaves\altaffilmark{23}}
\author{J. Gregson\altaffilmark{24, 25}}
\author{W. Holland\altaffilmark{26, 27}}
\author{G. Joncas\altaffilmark{28}}
\author{J.M. Kirk\altaffilmark{12}}
\author{L.B.G. Knee\altaffilmark{2}}
\author{S. Mairs\altaffilmark{1}}
\author{K. Marsh\altaffilmark{11}}
\author{B.C. Matthews\altaffilmark{2, 1}}
\author{G. Moriarty-Schieven\altaffilmark{2}}
\author{C. Mowat\altaffilmark{5}}
\author{S. Pezzuto\altaffilmark{29}}
\author{J. Rawlings\altaffilmark{30}}
\author{J. Richer\altaffilmark{7, 8}}
\author{D. Robertson\altaffilmark{31}}
\author{E. Rosolowsky\altaffilmark{32}}
\author{D. Rumble\altaffilmark{5}}
\author{N. Schneider-Bontemps\altaffilmark{33, 34}}
\author{H. Thomas\altaffilmark{3}}
\author{N. Tothill\altaffilmark{35}}
\author{S. Viti\altaffilmark{30}}
\author{G.J. White\altaffilmark{24, 25}}
\author{J. Wouterloot\altaffilmark{3}}
\author{J. Yates\altaffilmark{30}}
\author{M. Zhu\altaffilmark{36}}
\altaffiltext{1}{Department of Physics and Astronomy, University of Victoria, Victoria, BC, V8P 1A1, Canada}
\altaffiltext{2}{NRC Herzberg Astronomy and Astrophysics, 5071 West Saanich Rd, Victoria, BC, V9E 2E7, Canada}
\altaffiltext{3}{Joint Astronomy Centre, 660 N. A`oh\={o}k\={u} Place, University Park, Hilo, Hawaii 96720, USA}
\altaffiltext{4}{Max Planck Institute for Astronomy, K\"{o}nigstuhl 17, D-69117 Heidelberg, Germany}
\altaffiltext{5}{Physics and Astronomy, University of Exeter, Stocker Road, Exeter EX4 4QL, UK}
\altaffiltext{6}{Leiden Observatory, Leiden University, PO Box 9513, 2300 RA Leiden, The Netherlands}
\altaffiltext{7}{Astrophysics Group, Cavendish Laboratory, J J Thomson Avenue, Cambridge, CB3 0HE, UK}
\altaffiltext{8}{Kavli Institute for Cosmology, Institute of Astronomy, University of Cambridge, Madingley Road, Cambridge, CB3 0HA, UK}
\altaffiltext{9}{Department of Physics and Astronomy, University of Waterloo, Waterloo, Ontario, N2L 3G1, Canada  }
\altaffiltext{10}{LSST Project Office, 933 N. Cherry Ave, Tucson, AZ 85719, USA}
\altaffiltext{11}{School of Physics and Astronomy, Cardiff University, The Parade, Cardiff, CF24 3AA, UK}
\altaffiltext{12}{Jeremiah Horrocks Institute, University of Central Lancashire, Preston, Lancashire, PR1 2HE, UK}
\altaffiltext{13}{European Southern Observatory (ESO), Garching, Germany}
\altaffiltext{14}{Jodrell Bank Centre for Astrophysics, Alan Turing Building, School of Physics and Astronomy, University of Manchester, Oxford Road, Manchester, M13 9PL, UK}
\altaffiltext{15}{Current address: Institute for Astronomy, ETH Zurich, Wolfgang-Pauli-Strasse 27, CH-8093 Zurich, Switzerland}
\altaffiltext{16}{Universit\'e de Montr\'eal, Centre de Recherche en Astrophysique du Qu\'ebec et d\'epartement de physique, C.P. 6128, succ. centre-ville, Montr\'eal, QC, H3C 3J7, Canada}
\altaffiltext{17}{James Madison University, Harrisonburg, Virginia 22807, USA}
\altaffiltext{18}{School of Physics, Astronomy \& Mathematics, University of Hertfordshire, College Lane, Hatfield, HERTS AL10 9AB, UK}
\altaffiltext{19}{Astrophysics Research Institute, Liverpool John Moores University, Egerton Warf, Birkenhead, CH41 1LD, UK}
\altaffiltext{20}{Imperial College London, Blackett Laboratory, Prince Consort Rd, London SW7 2BB, UK}
\altaffiltext{21}{Dept of Physics \& Astronomy, University of Manitoba, Winnipeg, Manitoba, R3T 2N2 Canada}
\altaffiltext{22}{Dunlap Institute for Astronomy \& Astrophysics, University of Toronto, 50 St. George St., Toronto ON M5S 3H4 Canada}
\altaffiltext{23}{Physics \& Astronomy, University of St Andrews, North Haugh, St Andrews, Fife KY16 9SS, UK}
\altaffiltext{24}{Dept. of Physical Sciences, The Open University, Milton Keynes MK7 6AA, UK}
\altaffiltext{25}{The Rutherford Appleton Laboratory, Chilton, Didcot, OX11 0NL, UK.}
\altaffiltext{26}{UK Astronomy Technology Centre, Royal Observatory, Blackford Hill, Edinburgh EH9 3HJ, UK}
\altaffiltext{27}{Institute for Astronomy, Royal Observatory, University of Edinburgh, Blackford Hill, Edinburgh EH9 3HJ, UK}
\altaffiltext{28}{Centre de recherche en astrophysique du Qu\'ebec et D\'epartement de physique, de g\'enie physique et d'optique, Universit\'e Laval, 1045 avenue de la m\'edecine, Qu\'ebec, G1V 0A6, Canada}
\altaffiltext{29}{Istituto di Astrofisica e Planetologia Spaziali, via Fosso del Cavaliere 100, I-00133 Rome, Italy}
\altaffiltext{30}{Department of Physics and Astronomy, UCL, Gower St, London, WC1E 6BT, UK}
\altaffiltext{31}{Department of Physics and Astronomy, McMaster University, Hamilton, ON, L8S 4M1, Canada}
\altaffiltext{32}{Department of Physics, University of Alberta, Edmonton, AB T6G 2E1, Canada}
\altaffiltext{33}{LAB/OASU Bordeaux, CNRS, UMR5804, Floirac, France}
\altaffiltext{34}{I. Physik. Insitut, University of Cologne, Cologne, Germany}
\altaffiltext{35}{University of Western Sydney, Locked Bag 1797, Penrith NSW 2751, Australia}
\altaffiltext{36}{National Astronomical Observatory of China, 20A Datun Road, Chaoyang District, Beijing 100012, China}
%
%
\begin{abstract}
The dust emissivity spectral index, \textbeta, is a critical parameter for deriving the mass and temperature of star-forming structures, and consequently their gravitational stability. The \textbeta \ value is dependent on various dust grain properties, such as size, porosity, and surface composition, and is expected to vary as dust grains evolve. Here we present \textbeta, dust temperature, and optical depth maps of the star-forming clumps in the Perseus Molecular Cloud determined from fitting SEDs to combined \textit{Herschel} and JCMT observations in the 160 \textmu m, 250 \textmu m, 350 \textmu m, 500 \textmu m, and 850 \textmu m bands. Most of the derived \textbeta, and dust temperature values fall within the ranges of 1.0 - 2.7 and 8 - 20 K, respectively. In Perseus, we find the \textbeta \ distribution differs significantly from clump to clump, indicative of grain growth. Furthermore, we also see significant, localized \textbeta \ variations within individual clumps and find low \textbeta \ regions correlate with local temperature peaks, hinting at the possible origins of low \textbeta \ grains. Throughout Perseus, we also see indications of heating from B stars and embedded protostars, as well evidence of outflows shaping the local landscape.
\end{abstract}

%
%
\section{Introduction}
\label{sec:intro}


Thermal dust emission is an excellent tracer of cold, star-forming structures. When observed in wide bands, the spectral energy distributions (SEDs) of these structures are dominated by optically thin, thermal dust emission at far-infrared and sub-millimeter wavelengths. The column densities, and consequently the masses, of star-forming structures can thus be estimated from their dust emission by assuming a dust opacity, $\kappa_{\nu}$, and a dust temperature, $T_d$. The dust opacity at frequencies $\nu \lesssim 6$ THz, however, has a frequency dependency typically modelled as a power law, characterized by the emissivity spectral index, \textbeta \ (e.g., \citealt{Hildebrand1983}). Depending on the dust model, $\kappa_{\nu}$ can be uncertain by up to a factor of 3 - 7 at a given frequency between 0.3 - 3 THz (see Figure 5 in \citealt{Ossenkopf1994}). The ability to determine mass precisely, therefore, depends heavily on how well \textbeta \ and the normalization reference opacity, $\kappa_{\nu_0}$, can be constrained. 

Measuring \textbeta \ is difficult because it requires the SED shape of the dust emission to be well determined, especially when $T_d$ is not already known from independent measurements. The ability to determine the true SED shape can be particularly susceptible to noise in the data and temperature variations along the line of sight, resulting in erroneous measurements of \textbeta \ and $T_d$ (e.g., \citealt{Shetty2009_II}; \citealt{Shetty2009_I}; \citealt{Kelly2012}). For this reason, a fixed \textbeta \ value of $\sim 2$ has commonly been adopted in the literature, motivated by both observations of diffuse interstellar medium (ISM; e.g., \citealt{Hildebrand1983}) and grain emissivity models (e.g., \citealt{Draine1984}).

Given that \textbeta \ is an optical property of dust grains and can depend on various grain properties such as porosity, morphology, and surface composition, the \textbeta \ value of a dust population may be expected to change as the population evolves. Indeed, observations have shown that dust within protostellar disks have \textbeta \ $\simeq 1$ (e.g., \citealt{Beckwith1991}), which is substantially lower than the diffuse ISM value ($\sim 2$). Furthermore, a wide range of \textbeta \ values ($1 \lesssim$ \textbeta \space $\lesssim 3$) has been reported in many observations of star-forming regions on smaller scales ($\lesssim 2'$; e.g., \citealt{Shirley2005}, \citeyear{Shirley2011}; \citealt{Friesen2005}; \citealt{Kwon2009}; \citealt{Schnee2010}; \citealt{Arab2012}; \citealt{Chiang2012}), as well as a few lower resolution observations on larger scales of a cloud (e.g., \citealt{Dupac2003}; \citealt{PlanckCollaboration2011}). These results suggest that dust grains in star-forming regions can evolve significantly away from its state in the diffuse ISM, prior to being accreted onto protostellar disks.

Recent large far-infrared/sub-millimeter surveys of nearby star-forming clouds such as the \textit{Herschel} Gould Belt Survey (GBS; \citealt{Andre2010}) and the James Clerk Maxwell Telescope (JCMT) GBS (\citealt{Ward-Thompson2007}) have provided unprecedented views of star-forming regions at resolutions where dense cores and filaments are resolved ($\lesssim 0.5'$). Despite this advancement, only a few studies (e.g., \citealt{Sadavoy2013}; \citealt{Schnee2014}; \citealt{Sadavoy2016}) have attempted to map out \textbeta \ at resolutions similar to these surveys over large areas. Such \textbeta \ measurements can be very valuable in providing more accurate mass and temperature estimates, and consequently gravitational stability estimates, of the structures revealed in these surveys. While the \textit{Herschel} GBS multi-band data may seem ideal for making \textbeta \ measurements in cold star-forming regions at first, Sadavoy et al. (\citeyear{Sadavoy2013}) demonstrated that \textit{Herschel} data alone are insufficient and longer wavelength data (e.g., JCMT 850 \textmu m observations) are needed to provide sufficient constraints on \textbeta. 

Here we present the first results of the JCMT GBS 850 \textmu m observation with the Sub-millimetre Common User Bolometer Array 2 (SCUBA-2) towards the Perseus Molecular Cloud as a whole. Employing the technique developed by Sadavoy et al. (\citeyear{Sadavoy2013}) for combining the \textit{Herschel} and JCMT GBS data, we simultaneously derive maps of $T_d$, \textbeta, and optical depth at 300 \textmu m, $\tau_{300}$, in the Perseus star-forming clumps by fitting modified blackbody SEDs to the combined data. We investigate the robustness and the uncertainties of our SED fits, and perform a detailed analysis of the derived parameter maps in an attempt to understand the local environment. In particular, we characterize the observed variation in \textbeta \ and discuss the potential physical driver behind such evolution. 

In this paper, we describe the details of our observation and data reduction in Section \ref{sec:obs}, and our SED-fitting method in Section \ref{sec:sedFitting}. The results, including the parameter maps drawn from SED fits, are presented in Section \ref{sec:results}. We discuss the implication of our results in Section \ref{sec:discussion}, with regards to radiative heating (Section \ref{subsec:thermal_feedbacks}), outflow feedback (Section \ref{subsec:outflow_feedbacks}), and dust grain evolution (Section \ref{subsec:betaNClumpEvolution}). We summarize our conclusions in Section \ref{sec:conclusion}.

\section{Observations}
\label{sec:obs}

\subsection{JCMT: SCUBA-2 Data}
\label{subsec:SCUBA2Data}

\begin{deluxetable*}{lcclcc}
\tablecaption{Details of the observed PONG regions \label{table:pongCoord}}
\tablewidth{0pt}
\tablehead{
\colhead{Scan Name} & \colhead{RA} & \colhead{DEC} & \colhead{Clump} & \colhead{Weather Grade\tablenotemark{a}} &
\colhead{Number of Scans}
}
\startdata
B5        & 03:47:36.92 & +32:52:16.5  & B5             & 2                                                                 & 6                                                                  \\
IC348-E   & 03:44:23.05 & +32:01:56.1  & IC 348         & 1                                                                 & 4                                                                  \\
IC348-C   & 03:42:09.99 & +31:51:32.5  & IC 348         & 2                                                                 & 6                                                                  \\
B1        & 03:33:10.75 & +31:06:37.0  & B1             & 1                                                                 & 4                                                                  \\
NGC1333-N & 03:29:06.47 & +31:22:27.7  & NGC 1333       & 1                                                                 & 4                                                                  \\
NGC1333   & 03:28:59.18 & +31:17:22.0  & NGC 1333       & 2                                                                 & 1                                                                  \\
NGC1333-S & 03:28:39.67 & +30:53:32.6  & NGC 1333       & 2                                                                 & 6                                                                  \\
L1455-S   & 03:27:59.43 & +30:09:02.1  & L1455          & 1                                                                 & 4                                                                  \\
L1448-N   & 03:25:24.56 & +30:41:41.5  & L1448          & 1                                                                 & 4                                                                  \\
L1448-S   & 03:25:21.48 & +30:15:22.9  & L1451          & 2                                                                 & 6                                                                  
\enddata
\tablecomments{The names, center coordinates, targeted clump names, and the weather grades of the individual observations made with the PONG1800 scan pattern. Clumps in the Table are ordered from east to west.}
\tablenotetext{a}{The weather grades 1 and 2 correspond to the sky opacity measured at 225 GHz of $\tau_{225} < 0.05$ and $0.05 \leq \tau_{225} < 0.07$, respectively.}
\end{deluxetable*}

Wide-band 850 \textmu m observations of Perseus were taken with the SCUBA-2 instrument (\citealt{Holland2013}) on the JCMT as part of the JCMT GBS program (\citealt{Ward-Thompson2007}). We included observations that were taken in the SCUBA-2 science verification (S2SV) and the main SCUBA-2 campaign of the GBS program, i.e., in October 2011, and between July 2012  and February 2014, respectively. As with the rest of the survey, Perseus regions were individually mapped using a standard PONG1800 pattern (\citealt{Kackley2010}, \citealt{Holland2013}, \citealt{Bintley2014}) that covers a circular region $\sim30'$ in diameter.

Our observations covered the brightest star-forming clumps found in Perseus, namely B5, IC 348, B1, NGC 1333, L1455, L1448, and L1451, listed here in order of east to west. Table \ref{table:pongCoord} shows the names and center coordinates of the observed PONG1800 maps, along with the weather grades in which they were observed. Based on priority, each planned PONG target was observed either four times under very dry conditions (Grade 1; $\tau_{225} < 0.05$) or six times under slightly less dry conditions (Grade 2; $\tau_{225} = 0.05 - 0.07$) to reach the targeted survey depth of 5.4 mJy beam$^{-1}$ for 850 \textmu m. The `northern' PONG region of NGC 1333 is the only exception, containing one extra S2SV PONG900 map observed under Grade 2 weather. As with Sadavoy et al. (\citeyear{Sadavoy2013}), we adopted $14.2''$ as our effective Gaussian FWHM beamwidth for the 850 \textmu m data based on the two components of the SCUBA-2 beam obtained from measurements.

All SCUBA-2 data observed for the JCMT GBS program were reduced with the \textsc{makemap} task from the \emph{Starlink} SMURF package (Version 1.5.0; \citealt{Jenness2011}; \citealt{Chapin2013}) following the JCMT GBS Internal Release 1 (IR1) recipe, consistent with other JCMT GBS first-look papers (e.g., \citealt{Salji2015a}). The $^{12}$CO $J=3-2$ line contribution to the 850 \textmu m data is further removed with the JCMT Heterodyne Array Receiver Program (HARP) observations (see \citealt{Buckle2009}) as part of the IR1 reduction. Weather-dependent conversion factors calculated by Drabek et al. (\citeyear{Drabek2012}) were applied to the HARP data prior to the CO subtraction. For more details on the IR1 reduction and CO subtraction, see Sadavoy et al. (\citeyear{Sadavoy2013}) and Salji et al. (\citeyear{Salji2015a}).

\subsection{\textit{Herschel}: PACS and SPIRE Data}
\label{subsec:PACSnSPIRE}

The Perseus region was observed with the PACS (Photodetector Array Camera and Spectrometer; \citealt{Poglitsch2010}) instrument and the SPIRE instrument (Spectral and Photometric Imaging Receiver; \citealt{Griffin2010}; \citealt{Swinyard2010}) as part of the \textit{Herschel} GBS program (\citealt{Andre2005}; \citealt{Andre2010}; \citealt{Sadavoy2012}, \citeyear{Sadavoy2014}), simultaneously covering the 70 \textmu m, 160 \textmu m, 250 \textmu m, 350 \textmu m, and 500 \textmu m wavelengths using the fast ($60''/$s) parallel observing mode. The observation of the western and the eastern portions of Perseus took place in February 2010 and February 2011, respectively, covering a total area of $\sim10$ deg$^2$. We reduced our data with Version 10.0 of the \textit{Herschel} Interactive Processing Environment (HIPE; \citealt{Ott2010}) using modified scripts written by M. Sauvage (PACS) and P. Panuzzo (SPIRE) and PACS Calibration Set v56 and the SPIRE Calibration Tree 10.1. Version 20 of the \textsc{Scanamorphos} routine (\citealt{Roussel2013}) was used in addition to produce the final PACS maps. The final \textit{Herschel} maps have resolutions\footnote{The beams are actually more elongated in the shorter wavelength bands due to fast mapping speed. The resolution quoted here are the averaged values.}  of $8.4''$, $13.5''$, $18.2''$, $24.9''$, and $36.3''$ in order of shortest to longest wavelength. For more details on the \textit{Herschel} observations of Perseus, see Pezzuto et al. (\citeyear{Pezzuto2012}), Sadavoy (\citeyear{Sadavoy2013PhDT}), and Pezzuto et al. (2016, in prep.).

The SCUBA-2 data are spatially filtered to remove slow, time-varying noise common to all bolometers, such as atmospheric emission to which ground-based sub-millimeter observations are susceptible. For effective combinations, we filtered the \textit{Herschel} data using the SCUBA-2 map-maker \textsc{makemap} (\citealt{Jenness2011}; \citealt{Chapin2013}) following the method described by Sadavoy et al. (\citeyear{Sadavoy2013}) to match the spatial sensitivity of the \textit{Herschel} data to that of the SCUBA-2 observations. Given that any zero-point flux offsets in the \textit{Herschel} data are removed by the spatial filtering along with other large-scale emission, no offset correction was applied to our \textit{Herschel} data. Further details on the parameters used for filtering the \textit{Herschel} data are described in Chen (\citeyear{Chen2015MsT})

\section{SED-fitting}
\label{sec:sedFitting}

We modelled our dust spectral energy distributions as a modified blackbody function in the optically thin regime in the form of 

\begin{equation} \label{eq:modBBF}
I_{\nu} = \tau_{\nu_0} \left ( \nu  / \nu_0 \right ) ^{\beta} B_{\nu}(T_d)
\end{equation}
where $\tau_{\nu_0}$ is the optical depth at frequency $\nu_0$, $\beta$ is the dust emissivity power law index, and $B_{\nu}(T_d)$ is the blackbody function at the dust temperature $T_d$. By adopting a dust opacity value, $\kappa_{\nu_0}$, one can further derive the gas column densities as the following using the $\tau_{\nu_0}$ values:
\begin{equation} \label{eq:colDen}
N(H_2) = \frac{\tau_{\nu0}}{\mu m_{H_2} \kappa_{\nu0}}
\end{equation}
where $\mu$ is the mean molecular weight of the observed gas and $m_{H_2}$ is the mass of a molecular hydrogen in grams. For this study, we adopted a reference frequency $\nu_0 = 1$ THz (300 \textmu m), $\mu = 2.8$, and $\kappa_{\nu_0} = 0.1$ cm$^2$ g$^{-1}$, consistent with assumptions made by the \textit{Herschel} GBS papers (e.g., \citealt{Andre2010}). 

We convolved the CO-removed SCUBA-2 850 \textmu m data and the spatially-filtered \textit{Herschel} data \textbf{with Gaussian kernels} to a common resolution of $36.3''$ to match that of the 500 \textmu m \textit{Herschel} map, the lowest resolution of our data. We re-aligned and re-gridded the convolved maps to the original 500 \textmu m \textit{Herschel} map, which has $14''\times14''$ pixels. The 70 \textmu m data were excluded from our SED fitting because that emission may trace a population of very small dust grains that are not in thermal equilibrium with the dust traced by the longer wavelength emission (\citealt{Martin2012}). 

Following Sadavoy et al. (\citeyear{Sadavoy2013}), we applied color correction factors of 1.01, 1.02, 1.01, and 1.03 to the 160 \textmu m, 250 \textmu m, 350 \textmu m, and 500 \textmu m \textit{Herschel} data, respectively, to account for the spectral variation within each band. The correction factors are taken from the averaged values calculated from SED models with specific $T_d$ and \textbeta \ values (i.e., 10 K $\leq T_d \leq 25$ K and $1.5 \leq$ \textbeta $\leq 2.5$). The color \textbf{calibration} uncertainties associated with these color correction factors are 0.05, 0.008, 0.01, and 0.02 times that of the color corrected values, respectively. 

For each pixel of the map with signal-to-noise ratios (SNRs) $\geq 10$ in all five bands, we fitted a modified blackbody function (Equation \ref{eq:modBBF}) to get best estimates of \textbeta, $T_d$, and $\tau_{\nu_{0}}$. We employed the minimization of $\chi^2$ method using the \textit{optimize.curve\textunderscore fit} routine from the \textit{Python} \textit{SciPy} software package, which uses the Levenberg-Marquardt algorithm for minimization. The flux uncertainties were calculated as the quadrature sum of the color calibration uncertainties and the map sensitivities (see Table \ref{table:rmsNoiseConv}), and were adopted as standard errors for the $\chi^2$ calculation. For each clump, we adopted the 1-$\sigma$ rms noise measured from a relatively smooth, emission-free region within the \textbf{convolved,} filtered maps as the map sensitivity.

In addition to color \textbf{calibration} uncertainties and map noise, \textbf{we followed K{\"o}nyves et al. (\citeyear{Konyves2015}) in adopting 20\% as the absolute flux calibration uncertainties for the PACS 160 \textmu m band and 10 \% for the SPIRE bands. We also followed Sadavoy et al. (\citeyear{Sadavoy2013}) in adopting 10\% as the absolute flux calibration uncertainty in the SCUBA-2 850 \textmu m band. We note that 10\% is a conservative estimate of the absolute flux calibration uncertainties for extended sources and the calibration accuracy of point sources are much higher ($<  7\%$ for PACS bands,  \citealt{Balog2014}; $\sim 5\%$ for SPIRE bands, \citealt{Bendo2013}; $< 8\%$ for SCUBA-2 850 \textmu m band).}

\textbf{The flux calibration uncertainties are assumed to be correlated between the bands within an instrument, and we employed a simple Monte Carlo method to determine the most probable fit given the assumed calibration uncertainties based on a thousand instances.} The resulting probability distribution for a given pixel was collapsed onto the $T_d$ and \textbeta \ axes separately and fitted with a Gaussian distribution to determine the associated 1-$\sigma$ uncertainties of the derived $T_d$  and \textbeta. Due to the asymmetric distribution of the derived $\tau_{300}$, we used the best fit $\tau_{300}$ from another SED fit with $T_d$ and \textbeta \ fixed at their mean best-fit values from their respective Gaussian fits. We derived the uncertainty in $\tau_{300}$ from the covariance of this final fit. For more details on the SED fitting and how the flux calibration uncertainties are handled, see \citealt{Sadavoy2013} and \citealt{Chen2015MsT}. 

\textbf{The \textbeta \ and $T_d$ derived using the minimization of $\chi^2$ method can produce artificial anti-correlations due to the shape of the assumed SED becoming degenerate in the presence of moderate noise (\citealt{Shetty2009_I}; \citeyear{Shetty2009_II}) and flux calibration uncertainties. Detailed analysis of these correlated uncertainties are presented in Appendix \ref{apdx:betaT_antiCorr}. To ensure the robustness of our results, we rejected pixels with \textbeta \ uncertainties $>30\%$, where the fits to the SED are often poor and the \textbeta \ and $T_d$ uncertainties occupy a distinct part of the parameter space relative to the main population. We also removed isolated regions that contained less than 4 pixels and where Gaussian fits to the derived \textbeta \ and $T_d$ distributions from the Monte Carlo simulation were poor.}

\begin{deluxetable}{lccccc}
\tablecaption{Noise levels in the Perseus data \label{table:rmsNoiseConv}}
\tablewidth{0pt}
\tablehead{
\colhead{Band} & \colhead{160 \textmu m} & \colhead{250 \textmu m} & \colhead{350 \textmu m} & \colhead{500 \textmu m} &
\colhead{850 \textmu m}
}
\startdata
B5              & 60                     & 50                     & 30                     & 20                     & 20                     \\ 
IC 348           & 150                    & 110                    & 50                     & 20                     & 40                     \\ 
B1              & 80                     & 90                     & 60                     & 30                     & 30                       \\ 
NGC 1333        & 50                     & 60                     & 30                     & 20                     & 30                     \\ 
L1455           & 60                     & 70                     & 50                     & 20                     & 30                     \\ 
L1448           & 60                     & 50                     & 30                     & 20                     & 30                     \\
L1451           & 60                     & 50                     & 30                     & 20                     & 30                      
\enddata
\tablecomments{The approximate 1-$\sigma$ rms noise levels (mJy beam$^{-1}$) of the convolved, spatially filtered maps at the resolution of $36.3''$ for different Perseus clumps. The rms noise values were measured in a relatively smooth, emission-free region of the filtered clump maps.}
\end{deluxetable}

\section{Results}
\label{sec:results}

\subsection{Reduced Data}
\label{subsec:reducedData}

\begin{figure*}
\centering
\includegraphics[width=0.95\textwidth]{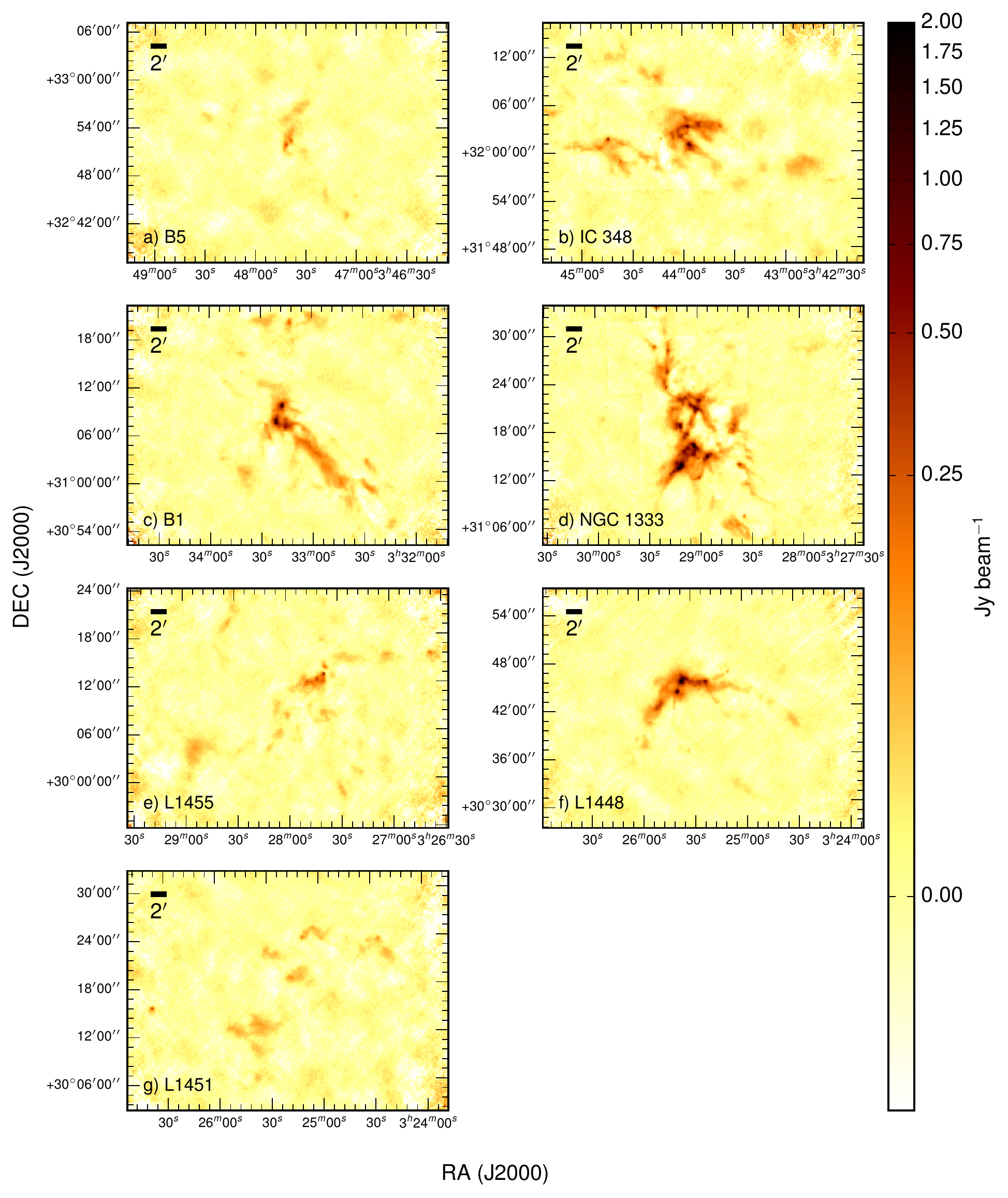}
\caption{SCUBA-2 850 \textmu m maps of the seven Perseus clumps, ordered from east to west and cropped to focus on the brightest regions. The flux are shown on a log scale from -0.03 Jy beam$^{-1}$ to 2.0 Jy beam$^{-1}$.}
  \label{fig:s2Map_Per}
\end{figure*}

\begin{figure*}
\centering
\includegraphics[width=0.95\textwidth]{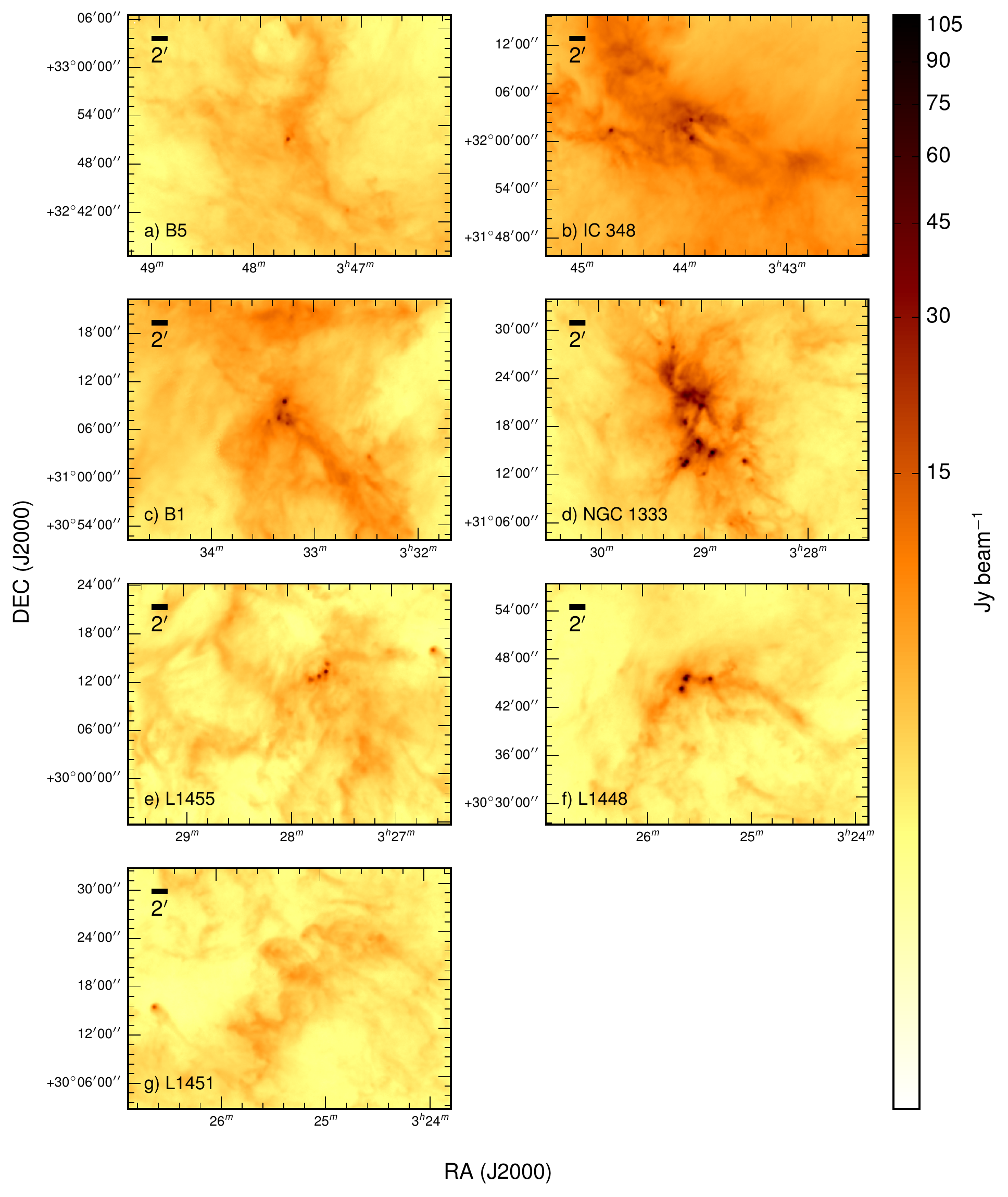}
\caption{Unfiltered SPIRE 250 \textmu m maps over the same regions shown in Figure \ref{fig:s2Map_Per}. The flux are shown on a log scale from 0.05 Jy beam$^{-1}$ to 110 Jy beam$^{-1}$.}
  \label{fig:her250Map_Per}
\end{figure*}

We present the reduced SCUBA-2 850 \textmu m maps of the B5, IC 348, B1, NGC 1333, L1455, L1448, and L1451 clumps in Figure \ref{fig:s2Map_Per}. \textbf{For a comparison, we also present the reduced, unfiltered SPIRE 250 \textmu m map in Figure \ref{fig:her250Map_Per}, cropped to match the same regions shown in Figure \ref{fig:s2Map_Per}. For the complete set of reduced, unfiltered PACS and SPIRE maps of Perseus, see Pezzuto et al. (2016, in prep.).}

\subsection{Derived Dust Temperatures}
\label{subsec:perseusTemp}

\begin{figure*}
\centering
\includegraphics[width=1.0\textwidth]{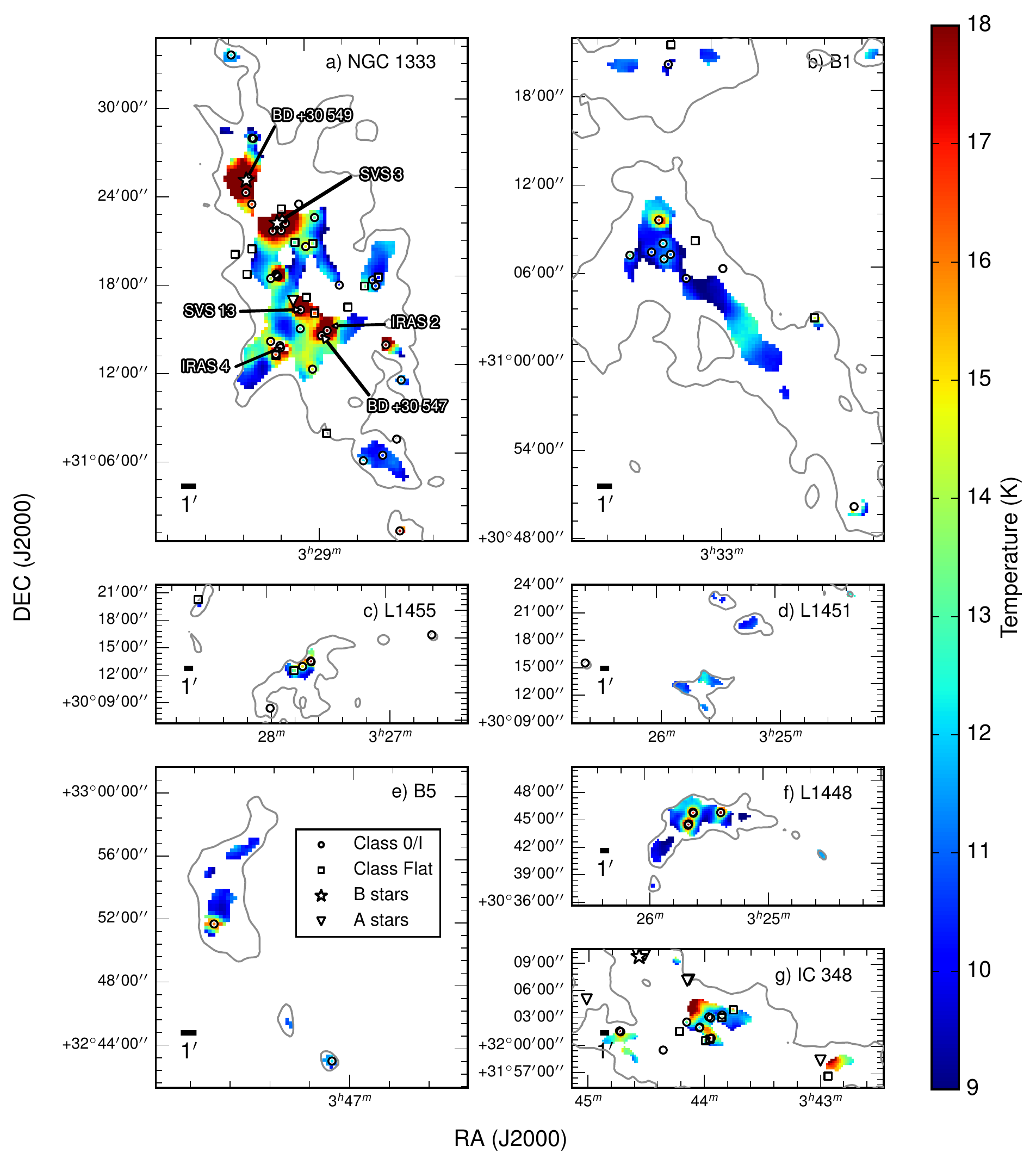}
\caption{Derived dust temperature maps of the seven Perseus clumps, with the colors scaled linearly from 9 K to 18 K. B and A stars in the region are denoted with star and triangle symbols, respectively. The Class 0/I and Flat YSOs identified in Dunham et al.'s Gould Belt catalogue (\citeyear{Dunham2015}) are denoted by circles and squares, respectively. Contours of the unfiltered Herschel 500 \textmu m emission at 1.5 Jy beam$^{-1}$ are overlaid on the maps in grey.} 
\label{fig:tempMap_Per}
\end{figure*}

Figure \ref{fig:tempMap_Per} shows the $T_d$ maps for all seven Perseus clumps overlaid with positions of embedded young stellar objects (YSOs), B stars, and A stars. The YSOs shown here are Class 0/I (circles) and Flat (squares) protostars identified from the Spitzer Gould Belt catalogue of mid-infrared point sources (\citealt{Dunham2015}). The B and A stars identified in various catalogs (referenced in the \textit{SIMBAD} database, \citealt{Wenger2000}) are labelled with star and triangle symbols, respectively. Contours of the unfiltered Herschel 500 \textmu m emission at 1.5 Jy beam$^{-1}$ are overlaid on the maps in grey. As demonstrated in Appendix \ref{apdx:betaT_antiCorr}, the uncertainty in the $T_d$ measurement due to flux calibration uncertainties is dependent on $T_d$. The typical uncertainties at $T_d$ values of $< 12$ K, 12 - 15 K, 15 - 20 K, and 20 - 30 K are 0.9 K, 1.5 K, 2.5 K, and 5.0 K respectively. The derived $T_d$ values at $T_d > 30$ K are poorly constrained and typically have uncertainties of $\gtrsim 10$ K. Our derived $T_d$ values are systematically lower than those obtained from unfiltered \textit{Herschel} data due to the preferential removal of the warm, diffuse dust emission by the spatial filtering. See Chen (\citeyear{Chen2015MsT}) for further details on the bias introduced by the spatial filtering.

The $T_d$ structures seen in Figure \ref{fig:tempMap_Per} often appear as circular, localized peaks overlaid on relatively cold backgrounds ($\sim 10-11$ K). Nearly all localized $T_d$ peaks are coincident with at least an embedded YSO or a B star along their respective lines of sight, indicating that these objects are likely the source of local heating. The few $T_d$ peaks that do not contain an embedded YSO, B star, or A star are all located in IC 348 and partially on the map edges. These $T_d$ peaks may be the result of local heating by sources outside of the map such as the nearby A star shown in Figure \ref{fig:tempMap_Per} or the nearby star cluster. 

The region just southeast of IRAS 2 in NGC 1333 also appears relatively warm ($\sim 14-15$ K) with no clear source of localized heating. While the star BD +30 547 (a.k.a., ASR 130, SSV 19) adjacent to IRAS 2B has often been classified as a foreground late type star (e.g., G2 IV, \citealt{Cernis1990}), Aspin (\citeyear{Aspin2003}) suggested that it may be a late-B V star with a cool faint companion, making BD +30 547 a potential source of heating in this region.

Interestingly, not all embedded YSOs are coincident with localized temperature peaks. This result is in agreement with that found by Hatchell et al. (\citeyear{Hatchell2013}) in NGC 1333.  Out of the 61 embedded YSOs located within the temperature maps of Perseus clumps, only seven are not coincident with local temperature peaks.  Some of these YSOs may be too faint, embedded, or young to have warmed their surrounding dust significantly. Alternatively, they could be more-evolved, less-embedded YSOs misidentified as Class 0/I or Class Flat objects, potentially due to line-of-sight confusion. 

\begin{figure}
\centering
\includegraphics[width=0.5\textwidth]{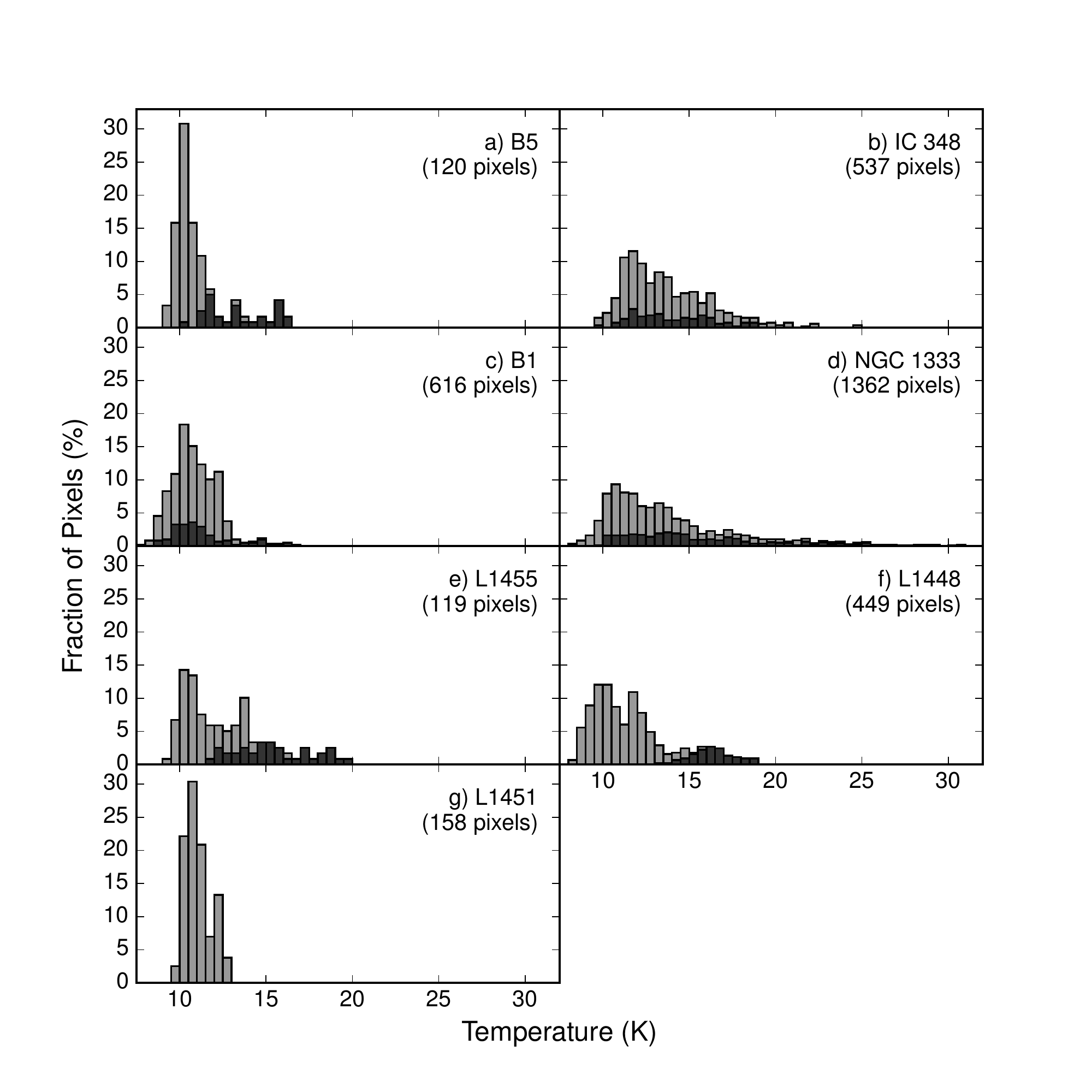}
\caption{Histograms of derived $T_d$ values in the seven Perseus clumps. All the pixels within a clump are shown in light grey while the pixels within a $72.6''$ diameter (i.e., two \textit{Herschel} 500 \textmu m beam widths) centered on a Class 0/I and Flat YSOs are shown in dark grey.}
  \label{fig:histo_temp_Per}
\end{figure}

Figure \ref{fig:histo_temp_Per} shows histograms of derived dust temperatures, $T_d$, in each of the Perseus clumps. The light and dark grey histograms represent all the pixels within a clump and the pixels found within a $72.6''$ diameter (i.e., two \textit{Herschel} 500 \textmu m beam widths) area centered on a Class 0/I and Flat YSOs, respectively. Globally, the $T_d$ distributions of Perseus clumps all share \textbf{two characteristic features:} a primary low $T_d$ peak and a high $T_d$ tail. \textbf{L1451 is the only clump without a high $T_d$ tail.} All clumps have their primary $T_d$ peaks located at $\sim10.5$ K, with the exception of IC 348 which has a temperature peak at $\sim11.5$ K. The former is consistent with the typical $T_d$ seen in NGC 1333 (\citealt{Hatchell2013}), the kinetic temperatures observed towards dense cores in Perseus with ammonia lines (\citealt{Rosolowsky2008}; \citealt{Schnee2009}), and the isothermal $T_d$ of prestellar cores in other clouds (\citealt{Evans2001}).

As shown in Figure \ref{fig:histo_temp_Per}, nearly all the pixels found in the high $T_d$ tails of the overall distribution (light grey) of B5, B1, L1455, and L1448 are located within a $72.6''$ diameter area centered on an embedded YSO (dark grey), indicating that these YSOs are indeed the main source of heating in these clumps. While protostellar heating also appears to be significant in IC 348 and NGC 1333, as seen in the $T_d$ maps (Figure \ref{fig:tempMap_Per}), only slightly less than half of the high $T_d$ pixels in these two clumps are found near embedded YSOs. The remainder of the pixels in the high $T_d$ tails of the distributions are likely from dust externally heated by B stars and nearby star clusters instead. \textbf{No embedded YSO is found near pixels in L1451 where the SEDs were fitted. This lack of local heating sources likely explains why L1451 does not have a high $T_d$ tail.}

\textbf{Looking from another prespective,} nearly all the pixels surrounding embedded YSOs (dark grey) in B5, L1455, and L1448 are found in the high $T_d$ tail of the overall distribution (light grey). This situation, however, is not seen in IC 348, B1, and NGC 1333, where a large fraction of the pixels near the embedded YSOs are found below $\sim 13$ K, near the 10.5 K $T_d$ peak. These low $T_d$ pixels are associated with the embedded YSOs not found towards the locally heated regions, or in some cases (e.g., B1) towards localized $T_d$ peaks that appear relatively cool.

Our derived $T_d$ values do not consistently drop off near map edges. This result indicates that our $T_d$ derivation is less susceptible to the edge effects caused by the large-scale filtering seen in the $T_d$ map of NGC 1333 derived by Hatchell et al. (\citeyear{Hatchell2013}) using the 450 \textmu m and 850 \textmu m data from the early SCUBA-2 shared risk observations. The north-western edge of our NGC 1333 map was the only exception, where the edge of the fitted map lies near the edge of the external mask used for the data reduction and spatial filtering. Nevertheless, the regions in Perseus where SEDs were fit are typically located well within the adopted external masks and are therefore unlikely to have experienced such a problem. We improved $T_d$ derivations in NGC 1333 with respect to Hatchell et al.'s analysis by using four additional bands from \textit{Herschel} and allowing \textbeta \ to vary. Furthermore, the JCMT GBS data used here were taken with longer integration time with the full SCUBA-2 array instead of just one sub-array and had less spatial filtering. At $T_d \sim 10$ K, Hatchell et al.'s $T_d$ uncertainties due to flux calibration alone is $\sim 1.2$ K, whereas our typical uncertainty is $\sim 0.9$ K.

The appearance of our derived $T_d$ map of B1 is morphologically consistent with that derived by Sadavoy et al. (\citeyear{Sadavoy2013}) using the same observations reduced with earlier recipes. In other words, the structures in the two $T_d$ maps are qualitatively the same in most places. Due to improvements in the data reduction, our $T_d$ values are more reliable than those derived by Sadavoy et al., and tend to be systematically colder by $\lesssim 1$ K at $T_d \lesssim 10$ K and warmer by $\lesssim 1$ K at $T_d \gtrsim 11$ K. The overall $T_d$ distribution of our map is broader near the 10.5 K $T_d$ peak than that of Sadavoy et al., with a more pronounced higher $T_d$ tail.

\subsection{Derived \textbeta \ and $\tau_{300}$}
\label{subsec:perseusBetaTau}

\begin{figure*}
\centering
\includegraphics[width=1.0\textwidth]{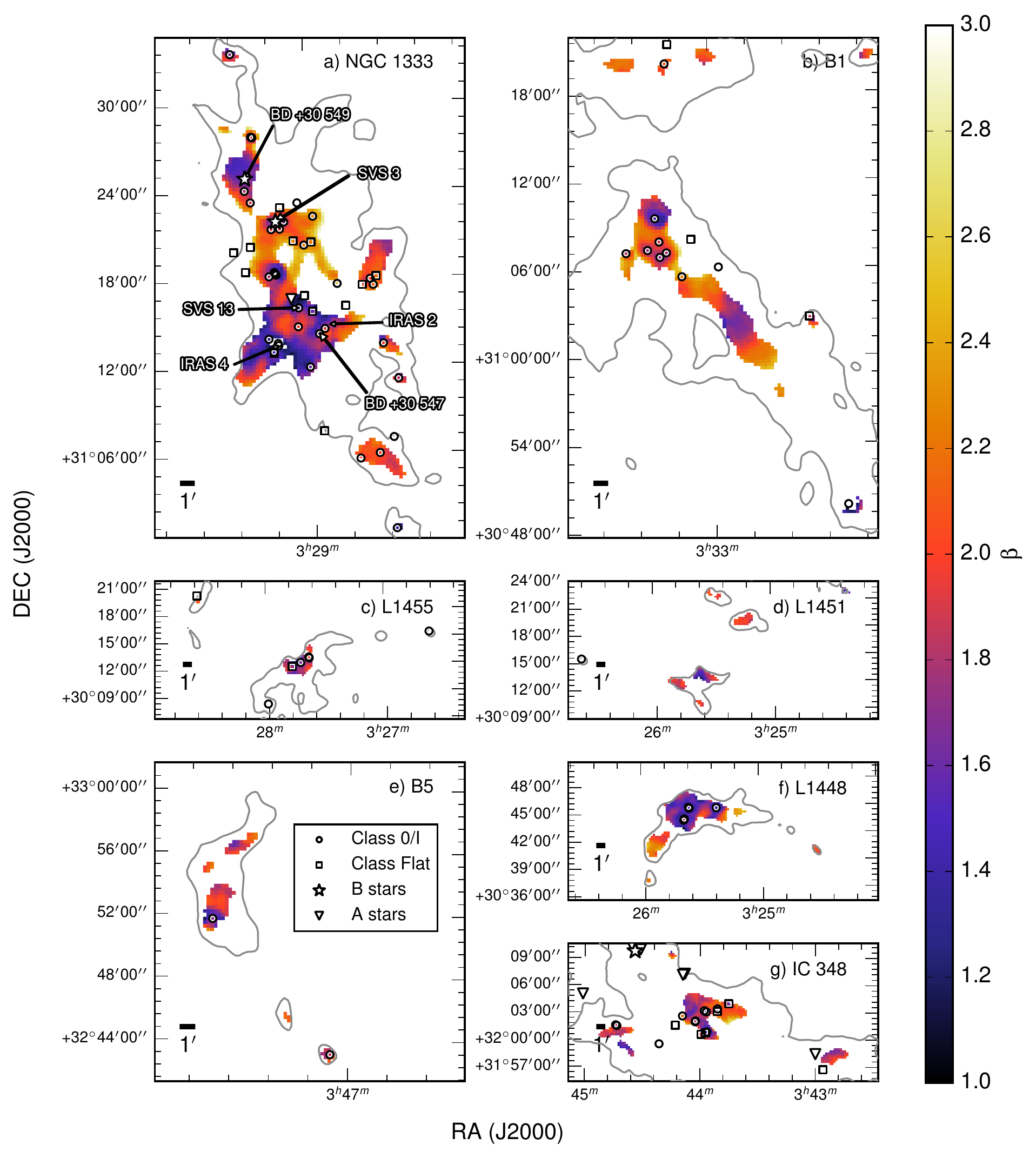}
\caption{Derived \textbeta \ maps of the seven Perseus clumps, with the colors scaled linearly from 1 to 3. Symbols and contours are the same as in Figure \ref{fig:tempMap_Per}.} \label{fig:betaMap_Per}
\end{figure*}

Figure \ref{fig:betaMap_Per} shows maps of derived \textbeta \space in the seven Perseus clumps. \textbf{The 1-$\sigma$ uncertainties in the \textbeta \ measurement due to flux calibration uncertainties are relatively independent of the derived \textbeta \ values and have a median value of $\sim 0.35$ in Perseus.} Pixels with similar \textbeta \space values tend to form well-defined structures, indicating that \textbeta \space variations seen in these clumps are correlated with local environment and not noise artifacts. Interestingly, low \textbeta \space structures tend to correlate with local $T_d$ peaks.

To a lesser extent, low \textbeta \ structures also appear to correlate with outflows traced by CO emission (e.g., see Figure \ref{fig:detailBetaTauOutflows} for details). The CO contribution to these maps has been subtracted although the percentage contamination seen in most pixels is less than the flux uncertainties of 10\% \textbf{with the exception of a few special, local cases} (\citealt{Chen2015MsT}). The additional uncertainty on the 850 \textmu m fluxes associated with the CO removal process is negligible. The resemblance between some of the \textbeta \space minima and outflow structures is therefore unlikely to be an artifact due to errors in the CO decontamination. While no study of free-free emission has been conducted over these outflows to assess whether or not such emission can be a significant contaminant in our data, free-free emission at centimetre wavelengths observed in radio jets is generally $<$ 1 mJy (\citealt{Anglada1996}). Free-free emission is also expected to be relatively weak at 850 \textmu m compared to the RMS noise of our 850 \textmu m data, given that these jets have widths much smaller than the JCMT beam. High angular resolution observations of the outflow sources SVS 13 (\citealt{Rodriguez1997}; \citealt{Bachiller1998}) and IRAS 4A (\citealt{Choi2011}) in NGC 1333 with the Very Large Array (VLA) have also shown that free-free emission is negligible at $\lambda \lesssim 3$ mm at these locations. The VLA survey of the Perseus protostars conducted by Tobin et al. (\citeyear{Tobin2016}) also found free-free emission near protostars to be faint relative to sub-millimeter dust emission and very compact with respect to our convolved $36.3''$ beam. Free-free emission is therefore unlikely to contaminate our data.

Pixels with \textbeta \ $\geq 3$ are found only in NGC 1333, mostly on the north-western edge where the filtering mask may be affecting the emission, as discussed in Section \ref{subsec:perseusTemp}. Therefore, we do not trust these steep \textbeta \ values. Low \textbeta \space values ($\lesssim1.5$), however, are generally well within the filtering mask, and are likely robust against the filtering systematics.

\begin{figure}
\centering
\includegraphics[width=0.5\textwidth]{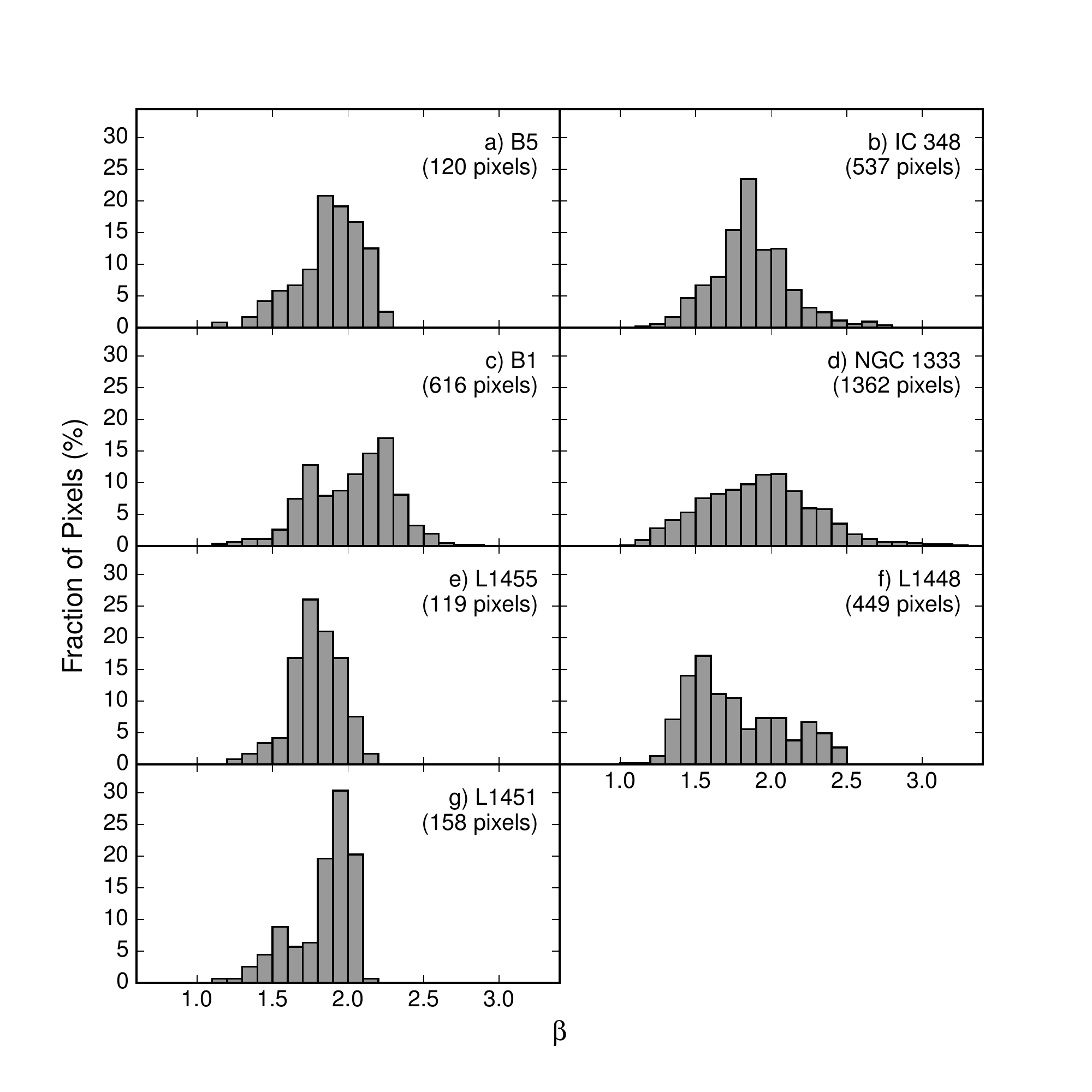}
\caption{Histograms of percentage of pixels with derived \textbeta \ values for seven Perseus clumps.}
  \label{fig:histo_beta_Per}
\end{figure}

Figure \ref{fig:histo_beta_Per} shows the histograms of derived \textbeta \space values from various clumps in Perseus. Unlike $T_d$, the \textbeta \space distributions of different clumps do not share a similar shape. While four clumps in Perseus appear to have a single peak (i.e., NGC 1333, IC 348, L1455, and B5), \textbf{the other three clumps (i.e., B1, L1448, and L1451) appear to have two or even three peaks.} The separations between \textbf{most of} these peaks are larger than the median \textbeta \ uncertainty ($\sim 0.35$), suggesting that these multiple peaks are not artifacts. The \textbeta \ values of the primary peak in each clump range between 1.5 (e.g., L1448) and 2.2 (e.g., B1), and the \textbeta \ values of most pixels range between 1.0 and 2.7. The latter range of \textbeta \ values is similar to that found in nearby star-forming clouds by Dupac et al. (\citeyear{Dupac2003}; $1.0 \leq$ \textbeta \space $\leq 2.5$) and for luminous infrared galaxies by Yang \& Phillips (\citealt{Yang2007}; $0.9 \leq$ \textbeta \space $\leq 2.4$).

The appearance of our derived \textbeta \ map of B1 is morphologically consistent with that derived by Sadavoy et al. (\citeyear{Sadavoy2013}) using the same observations reduced with earlier recipes. Due to improvements in the data reduction, our derived \textbeta \ values are more reliable. The majority of our derived \textbeta \ values are lower than those derived by Sadavoy et al. by $\lesssim 0.7$, with the remaining pixels having higher values by $\lesssim 0.3$. The overall distribution of our derived \textbeta \ is shifted downwards by $\sim 0.2$ relative to that of Sadavoy et al., with the primary peak being skewed towards the higher end of the distribution instead of the lower end.

\begin{figure*}
\centering
\includegraphics[width=1.0\textwidth]{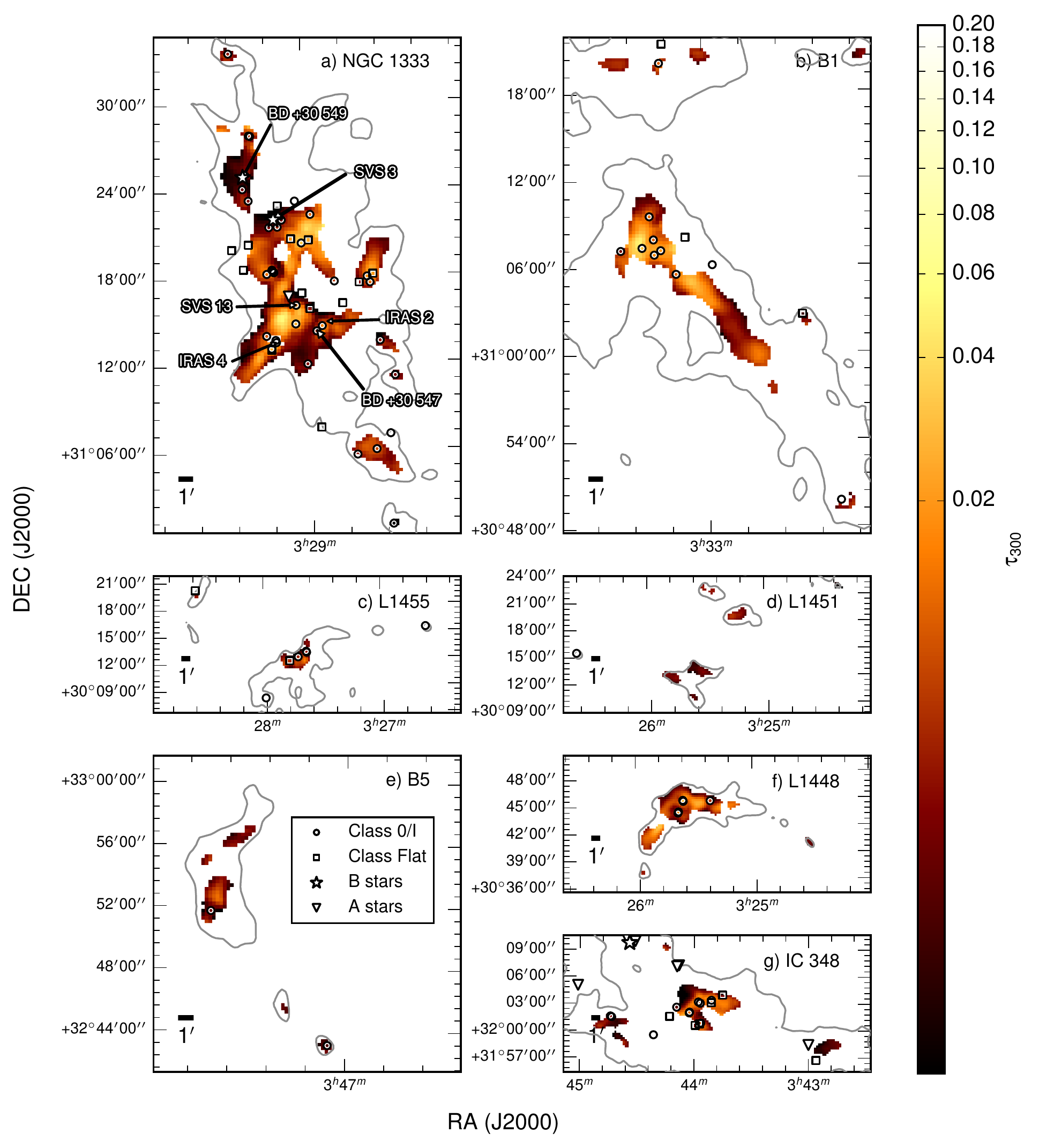}
\caption{Derived $\tau_{300}$ maps of the seven Perseus clumps, with the colors scaled logarithmically from 0.0035 to 0.1. Symbols and contours are the same as in Figure \ref{fig:tempMap_Per}.} 
\label{fig:tauMap_Per}
\end{figure*}

Figure \ref{fig:tauMap_Per} shows maps of derived $\tau_{300}$ in the seven Perseus clumps, overlaid with the same symbols as Figures \ref{fig:tempMap_Per} and \ref{fig:betaMap_Per}. Column densities can be further derived from $\tau_{300}$ using Equation \ref{eq:colDen} by assuming a $\kappa_{\nu_{0}}$ value. \textbf{The median uncertainty in the $\tau_{300}$ measurement derived from the covariance of the fit with $T_d$ and \textbeta \ fixed at the best determined values is $\sim 1.5 \%$. The typical uncertainties due to flux calibration uncertainties, however, as seen in the Monte Carlo simulation, is about a factor of 2 (see Appendix \ref{apdx:betaT_antiCorr}).} Higher $\tau_{300}$ structures seen in the B1, IC 348, L1448, and NGC 1333 clumps are found along filamentary structures, much like their dust emission counterparts. This similarity is less clear in the B5, L1455, and L1451 clumps where the areas where SEDs were fit (SNR $> 10$) are relatively small. 

Embedded YSOs are preferentially found towards local $\tau_{300}$ peaks. Many of these embedded YSOs, however, are spatially offset from the center of these peaks, often by slightly more than a beamwidth of our maps. The differences between \textbeta \ values found between these offsets are smaller than those expected from the anti-correlated uncertainties discussed in Appendix \ref{apdx:betaT_antiCorr}. Interestingly, we did not find any embedded YSOs towards the centers of the highest $\tau_{300}$ peaks in all Perseus clumps, with the exception of IC 348, which does not have $\tau_{300}$ peaks with distinct high values. The $T_d$ values of these starless $\tau_{300}$ peaks are also much lower than their surroundings, suggesting that these structures are cores well shielded from the interstellar radiation field (ISRF) due to their high column densities. 

\begin{deluxetable}{lccc}
\tablecaption{Derived column densities in Perseus clumps  \label{table:colDen}}
\tablewidth{0pt}
\tablehead{
\colhead{Clump} & \colhead{$\overline{N}\left (H_2 \right )$} & \colhead{$\sigma_{N \left (H_2 \right )}$} & \colhead{Number of pixels}
}
\startdata
B5              & 1.0                     & 0.54                      & 120                   \\ 
IC 348           & 1.4                    & 1.0                     & 537                    \\
B1              & 2.4                     & 1.8                     & 616                     \\
NGC 1333        & 1.9                     & 1.7                     & 1362                     \\
L1455           & 1.2                     & 0.59                     & 119                    \\
L1448           & 2.0                     & 1.3                     & 449                     \\            
L1451           & 0.78                    & 0.31                     & 158                              
\enddata
\tablecomments{The units of mean and standard deviation of column densities, $N(H_2)$, are both $10^{22}$ cm$^{-2}$.}
\end{deluxetable}

The $T_d$ values near embedded YSOs and towards higher $\tau_{300}$ regions also tend to be lower than in the lower $\tau_{300}$ regions, suggesting that protostellar heated regions can appear cooler due to being more deeply embedded and having more cool dust along the line of sight. Table \ref{table:colDen} shows the mean column densities found in each of the Perseus clumps. Indeed, the clumps with significant amounts of cold ($\lesssim 13$ K) pixels near embedded YSOs, i.e., B1, NGC 1333, and IC 348, have the first, third, and fourth highest mean column densities in Perseus, respectively. The B1 clump in particular has the majority of its pixels near embedded YSOs at $\lesssim 13$ K. 

\subsection{The \textbeta, $T_d$, and $\tau_{300}$ Relations}
\label{subsec:betaTempTauRelations}

\begin{figure}
\centering
\includegraphics[width=0.5\textwidth]{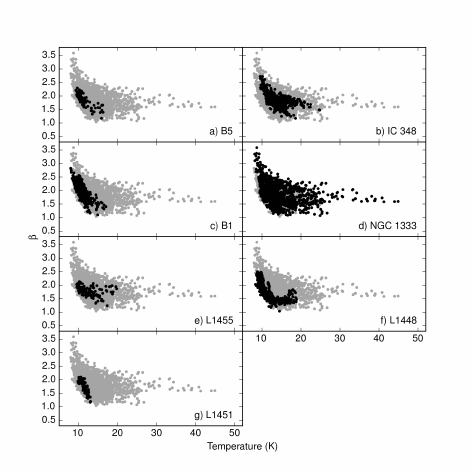}
\caption{Scatter plots of derived \textbeta \space vs. $T_d$ for each Perseus clump (black) overlaid onto the same for all Perseus clumps combined (grey).}
\label{fig:beta_vs_temp_Per}
\end{figure}

Figure \ref{fig:beta_vs_temp_Per} shows scatter plots of \textbeta \ versus $T_d$ for all seven Perseus clumps. At $T_d \lesssim 16$ K, anti-correlations between \textbeta \space and $T_d$ are found in all cases and cannot solely be accounted for by the anti-correlated \textbeta \space - $T_d$ uncertainties discussed in Appendix \ref{apdx:betaT_antiCorr}. At these $T_d$ values, there appears to be a prominent population of pixels in all seven clumps that exhibits a fairly linear relationship with a slope of $\sim -0.3$. Three clumps in particular, B1, B5, and L1451, seem to consist only of this population. \textbf{While this slope is similar to those found in the anti-correlated uncertainties presented in Appendix \ref{apdx:betaT_antiCorr} (see Figure \ref{fig:probCont_betaTemp_ngc1333}), the range of \textbeta \ found in this population is greater than the calculated \textbeta \ uncertainties.}

The other four Perseus clumps contain additional populations that extend into warmer temperatures ($T_d \gtrsim 16$ K) and have slopes which are much shallower than $-0.3$. In particular, the second population found in L1448 at $T_d \gtrsim 13.5$ K appears to have a constant \textbeta \ value of $\sim 1.4$ with a spread of $\pm \sim 0.15$. \textbf{This spread in \textbeta \ is smaller than the estimated uncertainties based on the assumed absolute flux calibration error, which is systematic across a map. This result suggests that our relative, pixel-to-pixel uncertainties within a map are indeed relatively small, and further indicates that there indeed is an underlying anti-correlation between \textbeta \ and $T_d$.} The \textbeta \ values found in these warmer populations are generally lower than those found at colder temperatures ($T_d \sim 10$ K), but are not necessarily the lowest values found within a clump.

\begin{figure}
\centering
\includegraphics[width=0.5\textwidth]{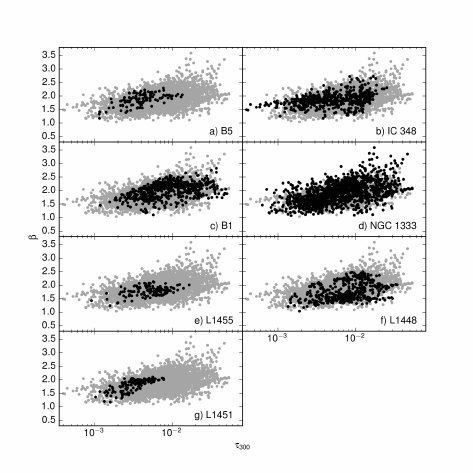}
\caption{Scatter plots of derived \textbeta \space vs. $\tau_{300}$ for each Perseus clump (black) overlaid onto the same for all Perseus clumps combined (grey).} 
\label{fig:beta_vs_tau300_Per}
\end{figure}

\textbf{Figure \ref{fig:beta_vs_tau300_Per} shows scatter plots of derived \textbeta \space versus $\tau_{300}$, where these appear to be weakly, postively correlated. Given the positive correlation between \textbeta \ and $\tau_{300}$ uncertainties discussed in Appendix \ref{apdx:betaT_antiCorr}, however, we do not consider this correlation to be significant.}

\section{Discussion}
\label{sec:discussion}

\subsection{Radiative Heating}
\label{subsec:thermal_feedbacks}

Throughout Perseus, we see evidence of embedded YSOs and nearby B stars heating their local environments. In contrast with the most common $T_d$ found in Perseus, $\sim$10.5 K, the $T_d$ near embedded YSOs are typically $\sim$ 14 - 20 K and can be well above 30 K near a B star where $T_d$ is not well constrained with our data. While not all embedded YSOs are associated with local $T_d$ peaks, all the local $T_d$ peaks coincide with locations of embedded YSOs and B stars. The only exceptions are the two extended warm regions in IC 348 that appear to be heated by sources outside of our parameter maps. Indeed, the main star cluster in IC 348 is located just northeast of the eastern warm region, while an A star is located just northeast of the western warm region. A higher local interstellar radiation field from the nearby young star cluster (\citealt{Tibbs2011}) likely explains why the most common $T_d$ found in IC 348 ($\sim$11.5 K) is slightly warmer than the other Perseus clumps.

The majority of the heated regions centered on embedded YSOs are larger than the FWHM beam of our map ($36.6''$), often having T$_{d} \geq 15$ K out to one full beam width from the center. In NGC 1333, all the locally heated regions with radii nearly one beam width in extent coincide with at least a Class 0 or I YSO with bolometric luminosity, $L_{bol}$, of 4 L$_\odot$ - 33 L$_\odot$ (\citealt{Enoch2009}). In other Perseus clumps, similar regions coincide with at least a Class 0 or I YSO with bolometric luminosity of 1 L$_\odot$ - 5 L$_\odot$. If we assume our distances to the western and eastern side of Perseus to be 220 pc (\citealt{Cernis1990}, \citealt{Hirota2008}) and 320 pc (\citealt{Herbig1998}), respectively, then these heated regions would have radii of $\sim 8000$ AU and $\sim 12000$ AU, respectively, which is somewhat less than the typical size of a dense core ($\sim $15000 AU in radius, \citealt{Andre2000}).

The core $T_d$ profile due to protostellar heating can be modelled with radiative transfer codes. Using \textit{Dusty} (\citealt{Ivezic1999}), J{\o}rgensen et al. (\citeyear{Jorgensen2006b}) modelled such $T_d$ profiles for various core density profiles, protostellar bolometric luminosities, and interstellar radiation fields (ISRF), assuming the OH5 dust opacities (\citealt{Ossenkopf1994}). For a core modelled after the MMS9 core in the Orion molecular cloud with density profile of $n = n_0(r/r_0)^{-1.5}$, where $r_0 = 50$ AU and $n_0 = 2.7 \times 10^8$ cm$^{-3}$, its $T_d$ drops down to 15 K at a radius of 2000 AU when an ISRF based on the solar neighborhood value (i.e., the standard ISRF) and a L$_{bol} =$ 10 L$_\odot$ for the central source are assumed. Despite having assumed an L$_{bol}$ that is relatively high, the size of this local heating is a factor of 4-6 smaller than what we find in Perseus. Given that the angular size of this modelled $T_d$ profile is smaller than our beam, further analysis that takes beam convolution \textbf{and its associated errors} into account will be needed to make a definitive comparison between the observed $T_d$ profile and the model.

Regions locally heated by B stars are more extended than regions heated by protostars. In NGC 1333, we find regions heated by B stars to be $\sim 1.6'$ in radius, corresponding to $\sim$ 21000 AU. Since these B stars are not embedded objects, the temperature profile of their surrounding dust can be reasonably approximated in the optically thin limit. An analytic expression of such a temperature profile first derived by Scoville \& Kwan (\citeyear{Scoville1976}) can be generalized as
\begin{equation} \label{eq:opThinTempPro}
T_{d}(r) = 50 Q_{abs}^{-q/2} \left ( \frac{r}{2\times10^{15}\textup{m}} \right )^{-q}\left ( \frac{L_{bol}}{10^5\textup{L}_{\odot}} \right ) \textup{K}
\end{equation}
where $Q_{abs}$ is the absorption efficient of a dust grain at $\lambda = 50$ \textmu m, and $q = 2/(4+\beta)$ (Equation 1, \citealt{Chandler1998}). For the expected range of \textbeta \ values between 1 and 2, the $q$ parameter does not vary significantly. The power-law slopes of the temperature profile at these \textbeta \ values resemble the \textit{Dusty} profile of J{\o}rgensen et al. (\citeyear{Jorgensen2006b}) at larger radii where the protostellar envelope is expected to be optically thin, before the ISRF heating starts to become dominant. 

We modelled the radial temperature profiles near the two B stars in NGC 1333 by assuming $\beta = 1.6$, the typical $\beta$ found near these two B stars, and normalizing Equation \ref{eq:opThinTempPro} to Wolfire \& Churchwell's (\citeyear{Wolfire1994}) dust emission models with $Q_{abs}^{-q/2} = 1.2$, as Chandler et al. (\citeyear{Chandler1998}; \citeyear{Chandler2000}) and J{\o}rgensen et al. (\citeyear{Jorgensen2006b}) did. Given that the B stars BD +30 549 and SVS 3 have luminosities of 42 L$_\odot$ (\citealt{Jennings1987}) and 138 L$_\odot$ (\citealt{Connelley2008}), respectively, the temperatures of these two stars are expected to drop down to 15 K at radii of $\sim$13000 AU and 24000 AU, comparable to the projected distances of 21000 AU (i.e., $\sim 1.6'$) that we observed. 

The localized, radiative heating due to YSOs and B stars may provide star-forming structures with additional pressure support. When treating a star-forming structure as a body of isothermal ideal gas with uniform density, the smallest length at which such an object can become gravitationally unstable due to perturbation, i.e., the Jeans length ($\lambda_{J}$), is related to the object's temperature by $\lambda_{J} \propto T^{1/2}$. The total mass enclosed within a Jeans length, i.e., the Jeans mass ($M_{J}$), scales as $M_{J} \propto T^{3/2}$.

Even though protostellar heating is the most common form of local heating observed in Perseus, protostellar cores are likely no longer Jeans stable given that they are currently collapsing into protostars. Protostellar heating could also provide additional Jeans stability to inter-core gas, but none of the warm regions in Perseus heated by solar-mass YSOs extends beyond the typical radius of a core ($\lesssim $ 15000 AU, \citealt{Andre2000}). 

While B stars are unlikely to have circumstellar envelopes themselves, they may be capable of heating nearby protostellar cores. Indeed, three SCUBA cores identified by Hatchell et al. (\citeyear{Hatchell2005}) are located in projection within these B-star heated regions: HRF57, HRF56, and HRF54. These cores do not contain known YSOs and have beam averaged $T_d$ between 19 K and 40 K. If we assume the densities of these cores remain unchanged in the process of heating and the gas in these cores are thermally coupled to the dust, then the associated Jeans masses of these cores at their respective temperatures are a factor of $\sim$ 3 - 8 higher than those of their counterparts at 10 K. 

\subsection{Outflow Feedback}
\label{subsec:outflow_feedbacks}

\begin{figure*}
	\centering
	\subfloat{\includegraphics[width=0.45\textwidth, trim = 0mm 9mm 0mm 18mm, clip = true]{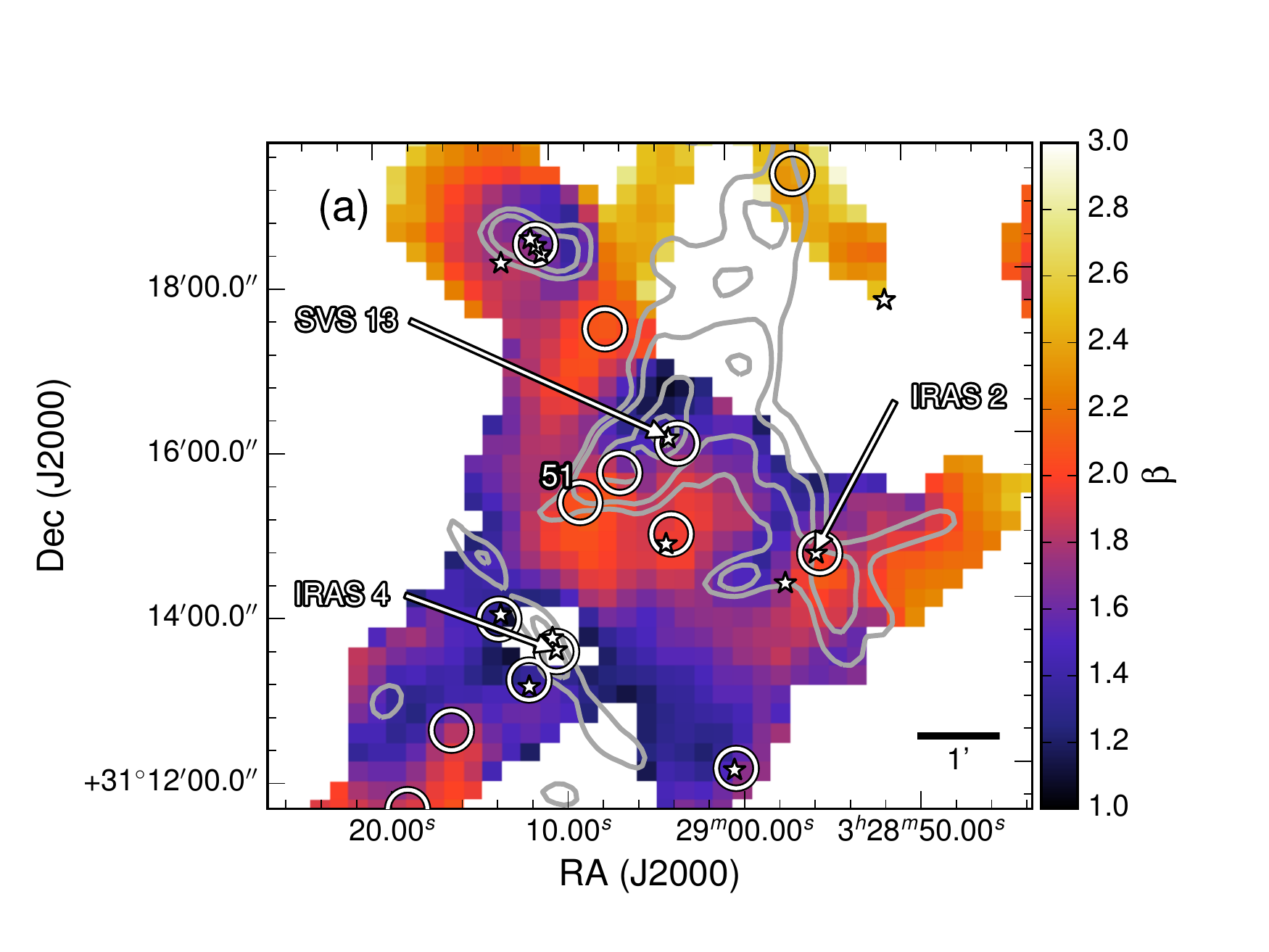}}
	\subfloat{\includegraphics[width=0.45\textwidth, trim = 0mm 9mm 0mm 18mm, clip = true]{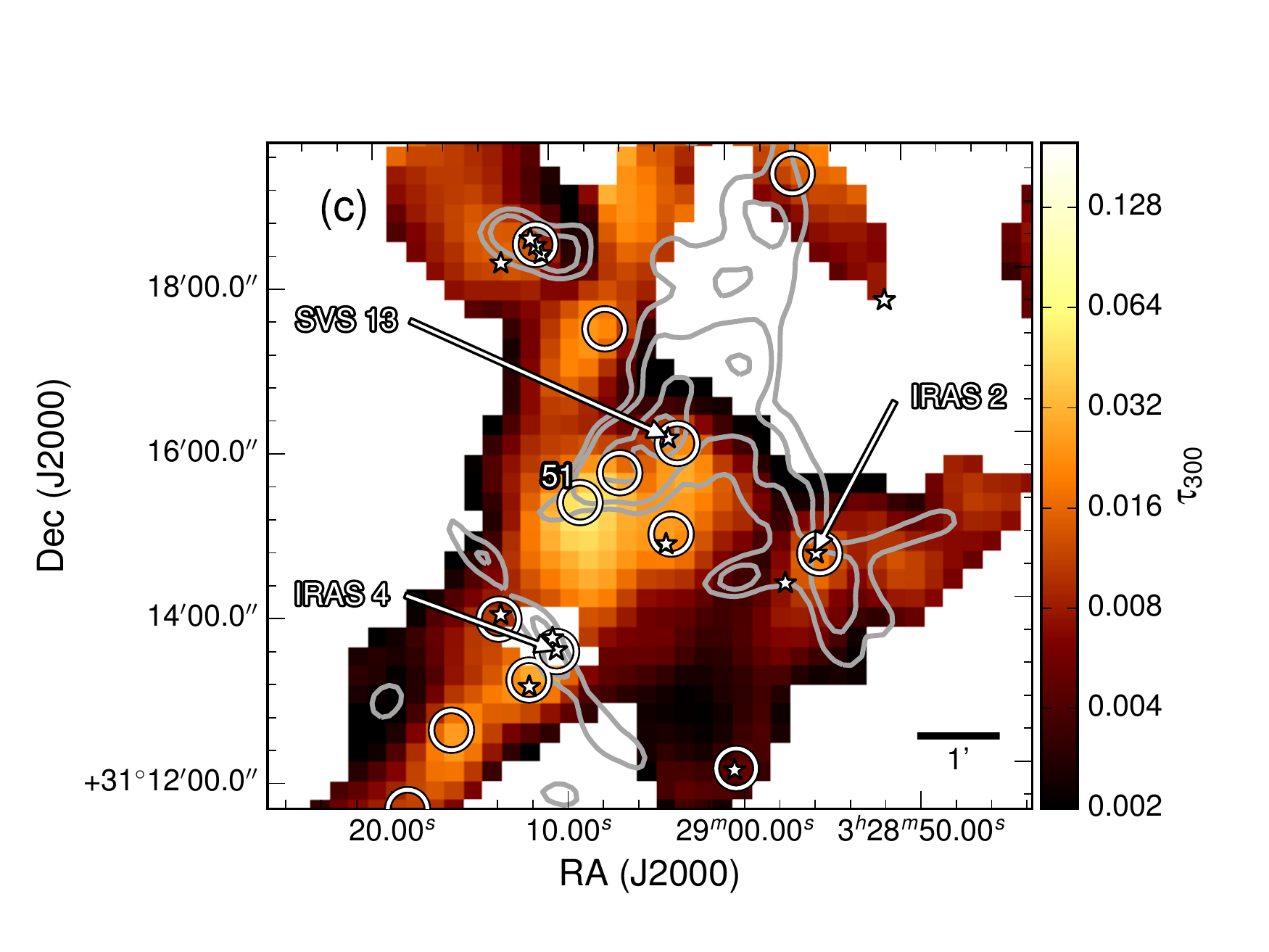}}
	\\
	\subfloat{\includegraphics[width=0.45\textwidth, trim = 0mm 9mm 0mm 18mm, clip = true]{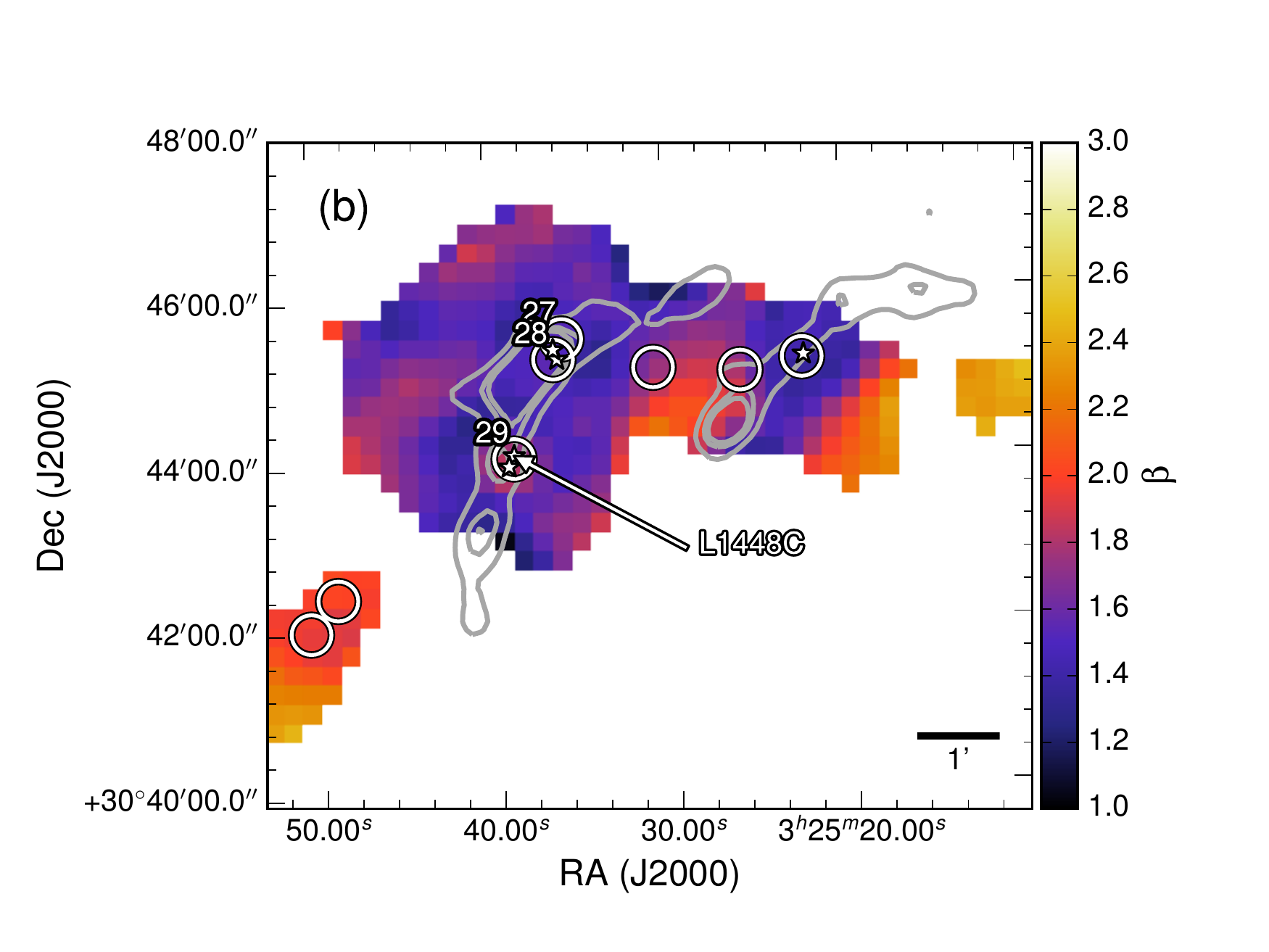}}
	\subfloat{\includegraphics[width=0.45\textwidth, trim = 0mm 9mm 0mm 18mm, clip = true]{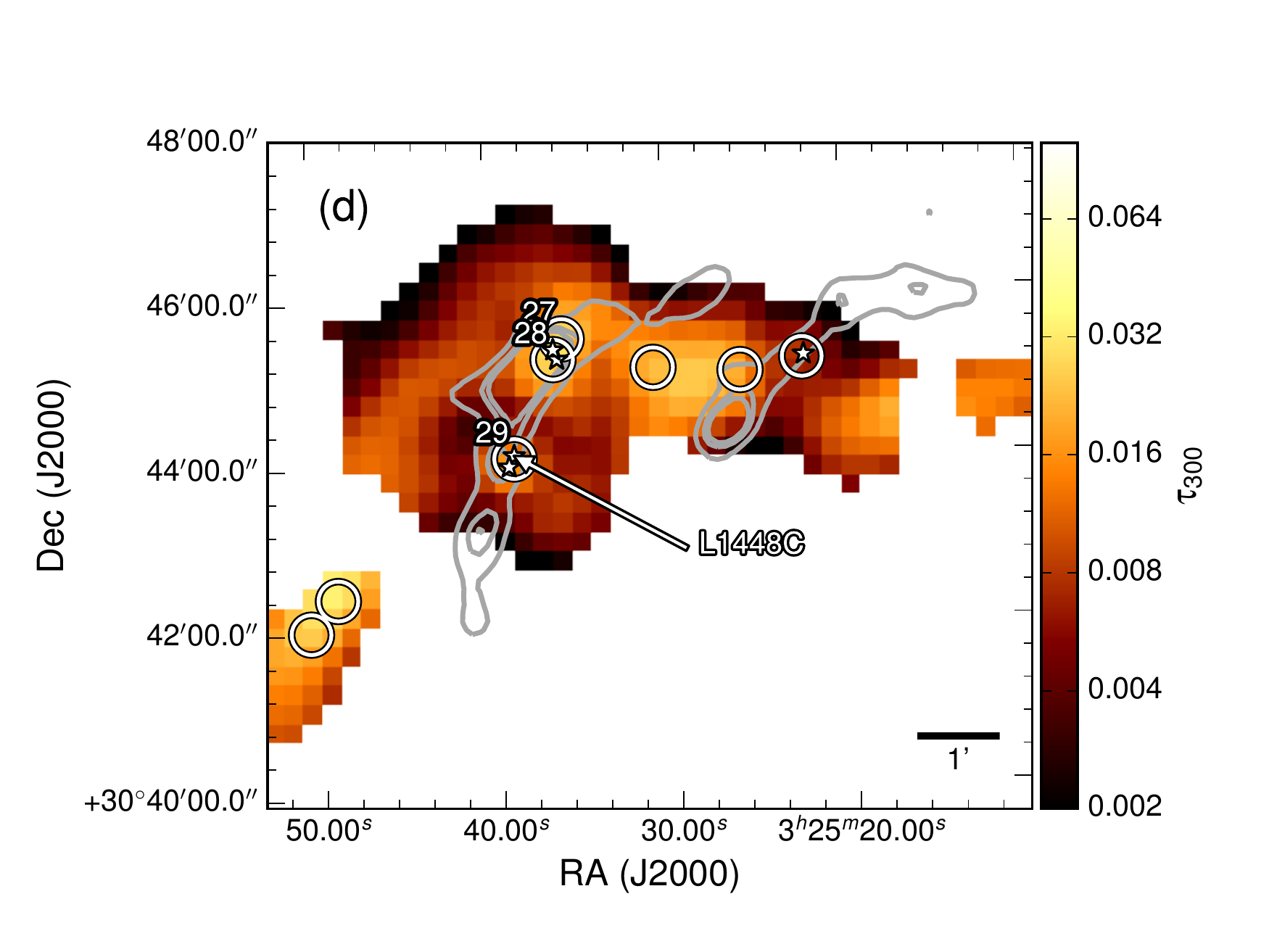}}
	\caption{Maps of derived \textbeta \ in a) NGC 1333 and b) L1448 and their $\tau_{300}$ counterparts c) and d), respectively. These maps are overlaid with the SCUBA cores identified by Hatchell et al. (\citeyear{Hatchell2005}; circles), Class 0/I YSOs identified by Dunham et al. (\citeyear{Dunham2015}; stars), and contours of spatially filtered, integrated $^{12}$CO 3-2 emission (grey). Contours of CO emission are drawn at 15 mJy beam$^{-1}$, 30 mJy beam$^{-1}$, 80 mJy beam$^{-1}$, and 110 mJy beam$^{-1}$ for the NGC 1333 maps and at 10 mJy beam$^{-1}$, 20 mJy beam$^{-1}$, and 25 mJy beam$^{-1}$ for the L1448 maps. The colors in panels a and b are scaled linearly between 1.0 - 3.0, and logarithmically between 0.002 - 0.2 and 0.002 - 0.1 for panels c and d, respectively. Only the cores mentioned in the text are identified with numbers designated by Hatchell et al.}
\label{fig:detailBetaTauOutflows}
\end{figure*}

Several studies of dust continuum emission in NGC 1333 (\citealt{Lefloch1998}; \citealt{Sandell2001}; \citealt{Quillen2005}), as well as other nearby star-forming clumps (e.g., \citealt{Moriarty-Schieven2006}), have suggested that outflows play a significant role in shaping the local structure of star-forming clumps. Indeed, kinematic studies of outflows in Perseus (e.g., \citealt{Arce2010}; \citealt{Plunkett2013}) showed that outflows can inject a significant amount of momentum and kinetic energy into their immediate surroundings. As with these prior dust continuum studies, we find several $\tau_{300}$ cavities that coincide with bipolar outflows, particularly towards the ends of outflow lobes, in NGC 1333 and L1448 (see Figures \ref{fig:detailBetaTauOutflows}). Due to warmer $T_d$ values, which can cause low column density dust to emit brightly, some of these cavities were not revealed directly by single-wavelength continuum emission alone. 

In addition to low $\tau_{300}$ regions, we also find two high $\tau_{300}$ regions in Perseus that may have been formed from compression by nearby outflows. For example, the high $\tau_{300}$ region near the starless core HRF51 in NGC 1333 has the highest $\tau_{300}$ values in the clump, and is also located towards the southeastern end of the outflow associated with the HH7-11 outflows driven by SVS 13 (\citealt{Snell1981}). The high $\tau_{300}$ region surrounding the HRF27 and HRF28 cores in L1448 also has the second highest $\tau_{300}$ values found in L1448, and is located on the northeastern end of the outflow originating from L1448C (HRF29, a.k.a. L1448-mm) near HH197 (e.g., \citealt{Bachiller1990a}; \citealt{Bally1997}). The presence of Herbig-Haro objects in these high $\tau_{300}$ regions indicates that outflows are indeed interacting with ambient gas at these locations and perhaps compressing the gas (and dust) in the process, resulting in the observed high $\tau_{300}$ values. Given that the $\tau_{300}$ values we find around the starless core HRF51 are unusually high in comparison with the rest of Perseus, HRF51 would be a good candidate for followup observations to test the limits of starless core stability. 

We do not find evidence for shock-heating from outflows in our $T_d$ maps, which is unsurprising given that shock heated gas is not well coupled to dust thermally (e.g., \citealt{Draine1983}; \citealt{Hollenbach1989}). The dominant local influence of outflows appears to be their ability to reshape local structures such as cores and filaments. Indeed, gas compression driven by outflows can stimulate turbulence to form starless cores beyond simple Jeans instability. For example, the extra material being pushed towards a starless core by outflows can provide the needed perturbation to trigger core collapse, as well as providing the core with extra mass to accrete. On the other hand, these outflows can in principle also inject turbulence into the structures they compress, providing additional pressure support for these dense structures to remain temporarily stable against Jeans instability. The mechanical feedback from outflows, therefore, is likely important in regulating star formation in clustered environments where it can enhance or impede the processes through compression or clearing, respectively.

\subsection{Beta and Clump Evolution}
\label{subsec:betaNClumpEvolution}

The global properties of a star-forming clump are expected to evolve significantly as star formation progresses within it. Given that dust grains can undergo growth inside cold, dense environments and potentially be fragmented or destroyed in energetic processes, dust grains themselves are expected to evolve over time. \textbf{If grain evolution in a clump is dominated by processes that result in a net change in \textbeta, such as growth in maximum grain sizes (\citealt{Testi2014}), then we should be able to observe the global values of \textbeta \ varying from clump to clump due to the different evolutionary stages of these clumps.}

\subsubsection{Beta Variation and its Relation to Temperature}
\label{subsubsec:betaVarDiscuss}

In Perseus, we find significant \textbeta \ variations within individual clumps. As shown in Figure \ref{fig:histo_beta_Per}, the majority of \textbeta \space values in these clumps range from 1.0 to 2.7. The spatial distribution of \textbeta \ is not erratic, and appears fairly structured on a local scale, suggesting that \textbeta \space is likely dependent on the properties of the local environment. Furthermore, the spatial transition between pixels with \textbeta \space $>2$ and $<1.6$, though relatively sharp, is not abrupt with respect to our beam width ($36.3''$ FWHM). \textbf{The overall appearance of the \textbeta \ structures, as well as their $T_d$ and $\tau_{300}$ counterparts, is fairly smooth and coherent on scales larger than the beam, suggesting that our derived \textbeta \ values are robust against random noise.}

In Figure \ref{fig:beta_vs_temp_Per}, our derived \textbeta \space and $T_d$ values appear to be anti-correlated, similar to behavior found in prior studies of nearby star-forming regions at lower angular resolutions (e.g., \citealt{Dupac2003}; \citealt{PlanckCollaboration2011}). As detailed in Appendix \ref{apdx:betaT_antiCorr}, this anti-correlation cannot be solely accounted for by the anti-correlated \textbeta \ - $T_d$ uncertainties associated with our SED fits. The fact that lower \textbeta \ regions tend to be spatially well structured and coincide with locations that are locally heated by known stellar sources further demonstrates that there indeed exists an anti-correlation between \textbeta \ and $T_d$ that is not an artifact of the noise.

Shetty et al. (\citeyear{Shetty2009_I}) have shown that line-of-sight $T_d$ variations can lead to an underestimation of \textbeta \space and an overestimation of density-weighted $T_d$ (column temperature) by fitting a single-component modified blackbody curve to the SEDs of various models of cold cores and warm envelopes that have a single \textbeta \space value. When these synthetic SEDs are noiseless and are equally well sampled in both the Wien's and Rayleigh-Jeans' regimes, Shetty et al. found that systematic exclusion of shorter wavelength data, i.e., the bands toward the Wien's regime, will allow the \textbeta \ and $T_d$ values drawn from fits to approach the true \textbeta \ values and column temperatures. A single-component modified blackbody curve simply does not fit well SEDs with multiple temperature components in the Wien's regime.

\begin{figure}
\centering
\includegraphics[width=0.5\textwidth]{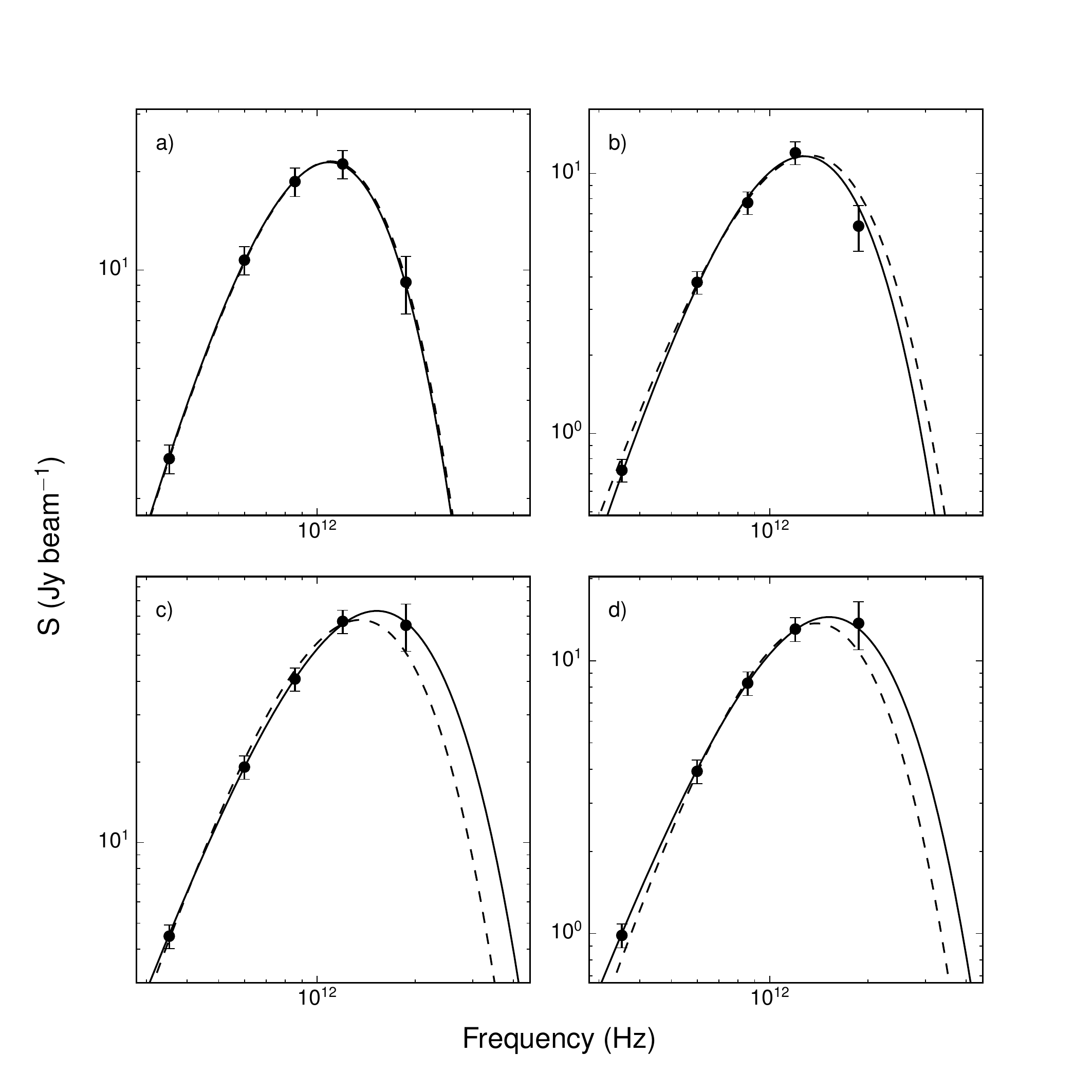}
\caption{Best SED fits to four selected pixels in NGC 1333 with $T_d$, \textbeta, and $\tau_{300}$ as free parameters (solid), and with \textbeta \ fixed at a value of 2 (dash). The best fit $T_d$ values in panels a, b, c, and d with \textbeta \ as a free parameter are 10.6 K, 11.4 K, 16.3 K, and 16.8 K, respectively. Their corresponding best fit \textbeta \ values are 2.0, 2.4, 1.6, and 1.4, respectively. The error bars represent the flux calibration uncertainties in each bands (i.e., 20\% for the SPIRE 160 \textmu m band and 10\% for the other bands).} 
\label{fig:sampSEDs}
\end{figure}

\textbf{In Perseus, we find the median $T_d$ of starless cores to be $9.8$ K when cores in IC 348 are excluded. The median $T_d$ of starless cores in IC 348 is a bit higher ($11.7$ K), likely due the clump having a higher ISRF. The former} $T_d$ is consistent with both the typical core $T_d$ found by Evans et al. (\citeyear{Evans2001}; $\sim10$ K) and the modelled column temperature of 9.6 K from Shetty et al. (\citeyear{Shetty2009_I}), which was itself based on the core density and $T_d$ profiles of Evans et al. The agreement between our derived $T_d$ and Shetty et al.'s modelled column temperature suggests that our fits to the SEDs of these cold cores are robust against line-of-sight $T_d$ variation. If the true \textbeta \ values along the lines of sight of our observations are fairly uniform, then our derived \textbeta \ values should approach those of the true \textbeta, as demonstrated by Shetty et al. Figure \ref{fig:sampSEDs}a shows a typical example of an SED found in cold core that is well fit by a single component modified blackbody curve.

In warmer regions, some of our SED fits have trouble reconciling the shortest wavelength band (i.e., 160 \textmu m), suggesting that line-of-sight $T_d$ variation may be a concern in these specific regions. Figures \ref{fig:sampSEDs}b-d show some examples. Given that preferential sampling towards the Rayleigh-Jeans tail will allow SED fits to place a tighter constraint on \textbeta, as Shetty et al. demonstrated, the \textbeta \ values we measured across 160 - 850 \textmu m toward the warmer regions should remain fairly robust against $T_d$ variation along the line of sight as the peaks of these SEDs shift further toward shorter wavelengths. Indeed, the fact that our sampled SED fits only have trouble fitting the 160 \textmu m bands in warmer regions suggests that these SED fits are preferentially anchored towards the longer wavelength points where \textbeta \ can be more robustly determined. Measurements of the flux ratio between the longer wavelength bands in NGC 1333 by Hatchell et al. (\citeyear{Hatchell2013}; SCUBA-2 450 \textmu m and 850 \textmu m bands) and in Perseus by Chen (\citeyear{Chen2015MsT}; SPIRE 500 \textmu m and SCUBA-2 850 \textmu m bands) have further confirmed that these regions are indeed warm and the $T_d$ drawn from the SED fits are not erroneous due to poor fits to the shorter wavelength bands.

\subsubsection{The Cause Behind \textbeta \ Variations}
\label{subsubsec:betaVarDiscuss}

\textbf{The \textbeta \ value of a grain population depends on various physical properties of its grains. The \textbeta \ value we derived in a given pixel should therefore be viewed as an effective \textbeta \ of the various underlying populations seen along the line of sight. If these effective \textbeta \ values are dominated by the \textbeta \ of a single grain population, then a simple comparison with laboratory measurements may reveal some physical properties of this particular population.}

Laboratory measurements have shown that \textbeta \space values can be intrinsically temperature dependent (e.g., \citealt{Agladze1996}; \citealt{Mennella1998}; \citealt{Boudet2005}). Many of these measured dependencies, however, are not significant enough to explain by themselves the anti-correlation seen here. Agladze et al. (\citeyear{Agladze1996}) found two species of amorphous grains with \textbeta \space values that decrease from $\sim$ 2.5 to $\sim$ 1.7 - 2 as the $T_d$ increases from 10 K to 25 K. While these measured temperature dependencies are not strong enough to be the sole cause of the anti-correlation seen here, they are perhaps significant enough to provide a partial contribution. 

Dust coagulation models have shown that grain growth can cause \textbeta \space values to decrease significantly from the typical diffuse ISM value (\textbeta \space $\sim$ 2), to as low as zero (e.g., \citealt{Miyake1993}; \citealt{Henning1995}). Since dust coagulation is favored in a denser environment, where the dust collision rates are higher, lower \textbeta \space values may be expected towards the densest regions.

As shown in Figure \ref{fig:beta_vs_tau300_Per}, we do not find evidence that \textbeta \space is anti-correlated with $\tau_{300}$, and thus column density. In contrast, we find that the highest $\tau_{300}$ structures in Perseus tend to have \textbeta \space $\gtrsim 2$, especially towards the colder regions not associated with localized heating. The lack of decrease in \textbeta \space towards these dense regions, however, is not sufficient to rule out grain growth through coagulation. The presence of ice mantles on the surface of dust grains, for example, can increase the \textbeta \ value from that of bare grains, i.e., $\sim 2$, to $\sim 3$ for dust typically found in diffuse ISM (\citealt{Aannestad1975}) and from $\sim 1$ to $\sim 2$ for modelled dust in the densest star-forming environments (\citealt{Ossenkopf1994}). Since the growth of ice mantles is also favored in the coldest and densest environments, such growth can, in principle, compete with coagulated grain growth by modifying \textbeta \space in opposite directions, causing the dust in some of the highest $\tau_{300}$ regions to have \textbeta \space $\sim 2$ instead of higher or lower values.  

Given that the presence of ice mantles on grain surfaces may be capable of suppressing the decrease in \textbeta \space caused by coagulated grain growth, we speculate that the \textbeta \space - $T_d$ anti-correlation may be explained by the sublimation of ice mantles in warmer environments. This hypothesis is useful in explaining why internally heated protostellar cores have lower \textbeta \space values than their starless, non-heated counterparts. Measuring the depletion levels of molecules of which ice mantles typically composed of towards heated and non-heated cores could test this hypothesis, provided these molecules can themselves survive in these heated environments. 

While lower \textbeta \ regions tend to coincide with locally heated regions, these \textbeta \ regions also tend to be more spatially extended than the heated regions. The low-\textbeta \ region surrounding IRAS 4\footnote{Many pixels immediately adjacent to IRAS 4 were removed due to their larger \textbeta \ uncertainties ($>30\%$).} in NGC 1333 is a prominent example of this behavior. Interestingly, many low \textbeta \space regions also coincide with the locations of bipolar outflows, prompting us to suspect that outflows may be capable of carrying lower \textbeta \space grains from deep within a protostellar core out to a scale large enough to be observed with our spatial resolution. Recent high resolution, interferometric studies on the inner regions of a small number of protostellar envelopes have found that \textbeta \ $\leq 1$ (\citealt{Chiang2012}; \citealt{Miotello2014}), suggesting that dense, inner regions of cores may indeed be a site for significant grain growth. Further high angular resolution observations of \textbeta \space within protostellar and prestellar cores may verify if low \textbeta \space grains are preferentially found deep within a dense core, and whether or not their production is associated with the presence of protostars or disks. 

Interestingly, the dependence of \textbeta \ on the maximum grain size, $a_{max}$, assuming a power-law grain size distribution, i.e., $n(a) \propto a ^{-q}$, behaves similarly regardless of other \textbeta \ dependencies (\citealt{Testi2014}; see their Figure 4). When $a_{max}$ is less than $\sim 10$ \textmu m, the \textbeta \ values computed by Testi et al. are all between $\sim$ 1.6 - 1.7, independent of the dust grain's chemical composition and porosity. As $a_{max}$ increases, \textbeta \ values for all three models first increase upwards to a maximum value of $\sim 2.5$ for compact grains, or $\sim 2$ for porous icy grains, and then quickly drop down to values less than 1 starting at $a_{max} \gtrsim 10^{3}$ \textmu m. The \textbeta \ values computed by Testi et al. are for wavelengths measured between 880 - 9000 \textmu m, which do not overlap with our observed wavelengths between 160 - 850 \textmu m. Nevertheless, given that a dust grain has less accessible modes to emit photons with wavelengths longer than or comparable to its physical size, we expect the same \textbeta \ dependency on $a_{max}$ to hold in our observed wavelength range with the corresponding $a_{max}$ shifted towards smaller values. Complementary \textbeta \ measurements at wavelengths greater than 850 \textmu m will test this idea.

If we assume the \textbeta \ dependency on $a_{max}$ to be the dominant, global driver of \textbeta \ evolution in a clump, as shown by Testi et al. (\citeyear{Testi2014}), while the other \textbeta \ dependencies are only locally significant, then \textbeta \ values observed on the global scale of a clump should be a good proxy to measure the relative evolutionary stage of a clump in terms grain growth. 

\subsubsection{Beta Variations Between Clumps}
\label{subsubsec:betaEvoClumps}

As demonstrated in Figure \ref{fig:histo_beta_Per}, the \textbeta \ distributions in Perseus differ significantly from clump to clump, suggesting that the dust grains are co-evolving with their host clumps. To make the measurement of \textbeta \ variation more sensitive, we binned our \textbeta \ values into three categories, with \textbeta \ values of 1.5 - 2, 2 - 3, and 0 - 1.5 to represent pristine, transitional, and evolved values of $a_{max}$, and thus \textbeta, respectively. This choice is based on Testi et al's (\citeyear{Testi2014}) modelled \textbeta \ - $a_{max}$ relations, and we extended the upper limit of the pristine, i.e., medium, \textbeta \ bin to a value of 2 to include the observed \textbeta \ values of $\sim 2$ in the diffuse ISM. While \textbeta \ values of dust do go through the ``pristine'' category again as they migrate from the transition category to the evolved category, this typically only occurs over a very narrow range of $a_{max}$ values relative to the overall $a_{max}$ values considered. To quantify the advancement of grain growth in each clump, we devise the growth index, $G$, as follows:
\begin{equation} \label{eq:growInd}
G = 100 - P_{pri} + P_{evo}
\end{equation}
where $P_{pri}$ and $P_{evo}$ are the percentage of pixels in each clump with the pristine and evolved values, respectively. As the typical $a_{max}$ in a clump increases, the $100 - P_{pri}$ term increases as \textbeta \ migrates away from the pristine values, resulting in an increase in $G$. When the grain growth enters an advanced stage, the increase in $a_{max}$ will also increase the value of $P_{evo}$, resulting in further increase in $G$.

\begin{figure}
\centering
\includegraphics[width=0.5\textwidth]{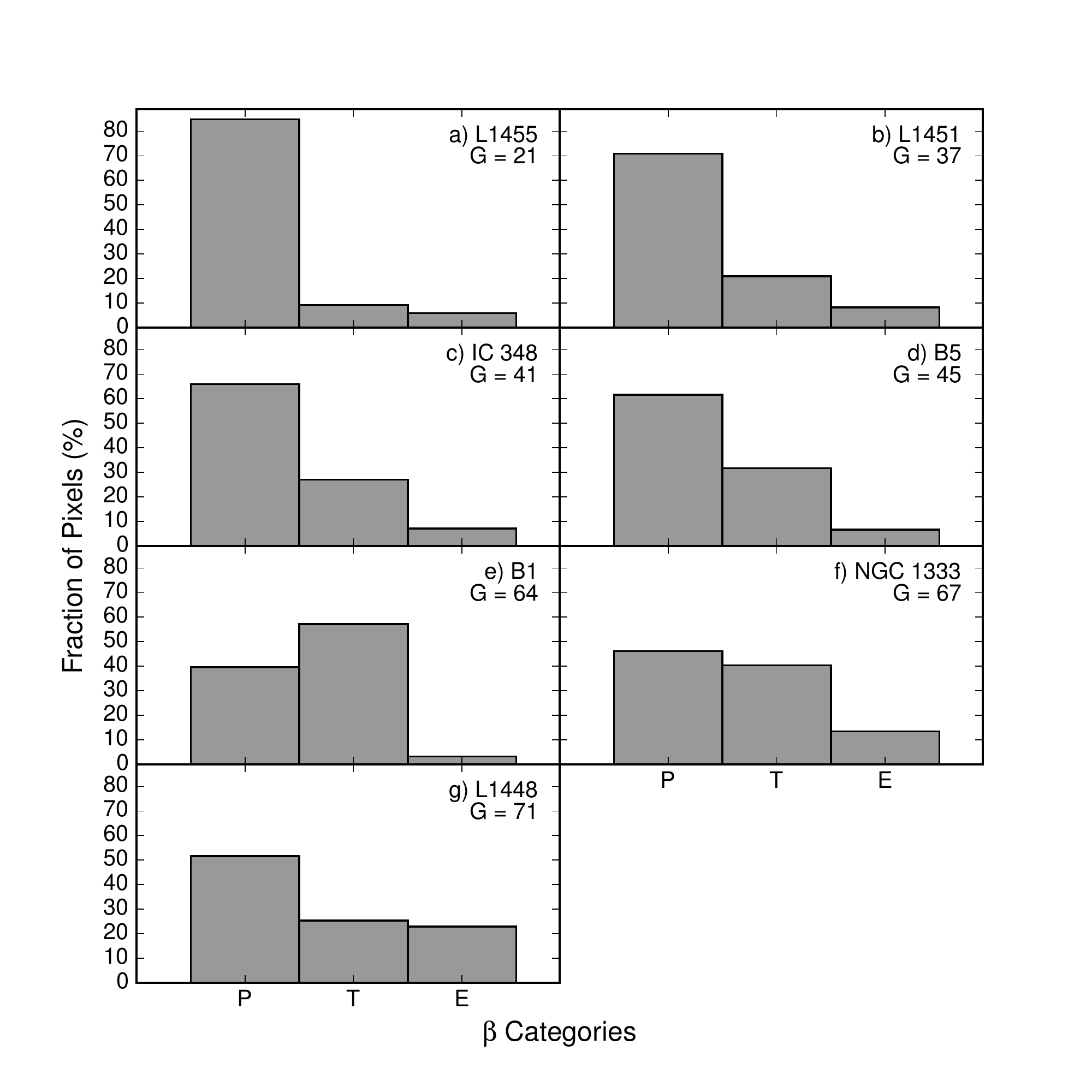}
\caption{Bar graph of derived \textbeta \ in the seven Perseus clumps binned to the three categories based on calculations by Testi et al. (\citeyear{Testi2014}): pristine (P), transitional (T), and evolved (E). The corresponding \textbeta \ values in these three categories are 1.5 - 2, 2 - 3, and 0 - 1.5, respectively. The order of the clumps are rearranged here from the least evolved to the most evolved as suggested by growth index, $G$, defined in Equation \ref{eq:growInd} and shown here in each panel.} 
\label{fig:beta_hist3bins}
\end{figure}

Figure \ref{fig:beta_hist3bins} shows the percentage of pixels in each of the \textbeta \ categories for the seven Perseus clumps and their corresponding $G$ values. The \textbeta \ bins are arranged in the order of pristine, transitional, and evolved categories from left to right. The order of the clumps here has been rearranged in the order of increasing $G$, corresponding to the perceived increase in $a_{max}$ based on calculations by Testi et al. (\citeyear{Testi2014}). As seen in Figure \ref{fig:beta_hist3bins}, the dust population in each of the Perseus clumps appears to be in a different evolutionary stage. \textbf{For a reference, the median \textbeta \ uncertainty is $\sim0.35$}. In terms of their relative dust evolution, the Perseus clumps are ordered as L1455, L1451, IC 348, B5, B1, NGC 1333, and L1448. We note that our sample sizes vary significantly from clump to clump, ranging from 1362 pixels in NGC 1333 to only 119 pixels in L1455. Our ability to make comparison between clumps is therefore limited by these sample size differences and the associated uncertainties in measuring \textbeta.  

\textbf{It may seem surprising at first that the measured \textbeta \ values would suggest IC 348 to be a less-evolved clump in Perseus}, given that IC 348 has often been considered to be an older star-forming clump near the end of its star-forming phase (\citealt{Bally2008}). The region \textbf{covered by} our SED fits, however, actually traces a filamentary ridge of higher density gas just southwest of the main optical cluster. This sub-region hosts a higher fraction of Class 0 and I YSOs and a lower fraction of Class II and III YSOs relative to the rest of IC 348 (\citealt{Herbst2008}). The relatively higher abundance of embedded YSOs in comparison to their more-evolved counterparts in this area of IC 348 suggests that this sub-region is indeed quite young. Due to low number statistics ($\leq 10$ YSOs), \textbf{we did not perform a similar consistency check on the youngest clumps, L1455 and L1451.}

L1448, NGC 1333, and B1 are likely the more-evolved clumps in Perseus based on their $G$ values. Indeed, the higher fractions of Class II and III YSOs relative to Class 0, I, and Flat YSOs seen in NGC 1333 (\citealt{Sadavoy2013PhDT}) suggest that NGC 1333 is one of the most evolved clumps in Perseus. Interestingly, B1 has the highest fraction of transitional \textbeta \ values and the lowest fraction of evolved \textbeta \ values in Perseus. Given that B1 also has the highest mean column density in Perseus (see Table \ref{table:colDen}), B1 may potentially be a younger clump with an enhanced grain-growth due to its higher density. 

In general, we do not expect the fraction of embedded YSOs and the relative dust evolutionary state to be tightly correlated given that the fraction of embedded YSOs is more indicative of the current star-forming rate whereas the dust evolutionary state measures the accumulated grain growth throughout a clump's history. Considering that grain growth is subjected to factors such as density and temperature over time, which may also differ between clumps, we do not expect the relative evolution we inferred from $a_{max}$ to be strictly proportional to time either. 

As mentioned in Section \ref{subsec:betaTempTauRelations}, we find a significant anti-correlation between \textbeta \ and $T_d$ in all seven Perseus clumps that cannot be solely accounted for by the anti-correlated uncertainties driven by noise or $T_d$ variation along the line of sight. Interestingly, there appears to be a population of pixels in all seven clumps that are located at the bottom left corner of the \textbeta \space - $T_d$ scatter plots with $T_d$ less than 20 K (see Fig. \ref{fig:beta_vs_temp_Per}). The anti-correlation displayed by this particular population appears fairly linear, and has a slope of $\sim -0.3$. This population is primarily responsible for the wide range of \textbeta \space values observed in each clump and its presence in all seven clumps may suggest a universal dust evolution that is common to all star-forming clumps. The \textbeta \ - $\tau_{300}$ distributions of Perseus clumps (see Fig. \ref{fig:beta_vs_tau300_Per}) all behave fairly similarly and do not show obvious trends associated with clump evolution.

\section{Conclusion}
\label{sec:conclusion}

In this study, we fit modified blackbody SEDs to combined \textit{Herschel} and JCMT continuum observations of star-forming clumps in the Perseus molecular cloud and simultaneously derived dust temperature, $T_d$, spectral emissivity index, \textbeta, and optical depth at 300 \textmu m, $\tau_{300}$, using the technique developed by Sadavoy et al. (\citeyear{Sadavoy2013}). We performed a detailed analysis of the derived $T_d$, \textbeta, and $\tau_{300}$ values in seven Perseus clumps, i.e., B5, IC 348, B1, NGC 1333, L1455, L1448, and L1451, and investigated anti-correlated uncertainties between \textbeta \ and $T_d$. We found that the \textbeta \ - $T_d$ anti-correlations seen in Perseus are significant and are not artifacts of noise or calibration errors. In addition, we demonstrated that our final SED fits are robust against systematic uncertainties associated with temperature variation along lines-of-sight.

We summarize our main findings as follows:

\begin{enumerate}
\item {The most common (i.e., mode) $T_d$ seen in Perseus clumps is $\sim 10.5$ K, except in IC 348 where it is $\sim 11.5$ K. The former $T_d$ is consistent with the kinetic temperatures of dense cores observed with ammonia emission lines in Perseus ($\sim$11 K; \citealt{Rosolowsky2008}; \citealt{Schnee2009}), and the isothermal $T_d$ of prestellar cores seen in various clouds ($\sim$10 K; \citealt{Evans2001}). The IC 348 clump being slightly warmer than the rest of Perseus may be due to stellar heating from the nearby young star cluster.} 

\item{Nearly all the local $T_d$ peaks found in Perseus coincide with locations of embedded YSOs (and nearby B stars in the case of NGC 1333), which indicates local heating by these YSOs. The immediate region surrounding an embedded YSO can have $T_d$ of up to $\sim 20$ K whereas that near a B star can have $T_d$ well above $\sim 30$ K.}

\item{We found significant \textbeta \space variations over individual star-forming clumps. Most \textbeta \ values found in Perseus are in the range of $1.0 \lesssim$ \textbeta \space$\lesssim 2.7$, similar to that found in nearby star-forming clouds by Dupac et al. (\citeyear{Dupac2003}; $1.0 \leq$ \textbeta \space $\leq 2.5$) and for luminous infrared galaxies by Yang \& Phillips (\citeyear{Yang2007}; $0.9 \leq$ \textbeta \space $\leq 2.4$). Maps of \textbeta, in general, appear well structured on a local scale,  indicating that \textbeta \space is likely a function of its local environment.}   
 
\item{We found \textbeta \space and $T_d$ to be anti-correlated in all Perseus clumps, and that lower \textbeta \space regions tend to coincide with local $T_d$ peaks. These anti-correlations cannot be solely accounted for by anti-correlated \textbeta \space and $T_d$ uncertainties associated with our SED fitting. The dust grains' intrinsic \textbeta \space dependency on temperature (e.g., \citealt{Agladze1996}) may be partially responsible for the anti-correlation, but is not strong enough to be the sole cause. The sublimation of surface ice mantles, which can increase \textbeta \space when present on a dust grain (e.g., \citealt{Aannestad1975}; \citealt{Ossenkopf1994}), may provide an explanation for the observed anti-correlation.} 

\item{Similar to prior studies (e.g., \citealt{Sandell2001}), we found cavities and peaks in maps of derived $\tau_{300}$, and thus column density, that coincide with the ends of some outflow lobes, suggesting that these structures were cleared out or compressed by local outflows, respectively. The fact that Herbig-Haro objects were found in locations where outflows meet the highest column density structures in individual clumps further supports this view. Creating column density maps from fitting SEDs with $T_d$, \textbeta, and $\tau$, is important for this analysis.} 

\item{The \textbeta \space distribution found in Perseus differs from one clump to another (see Figure \ref{fig:histo_beta_Per}). Binning \textbeta \ values into pristine (medium), transitional (high), and evolved (low) values of \textbeta \ as a function of maximum grain size, $a_{max}$, calculated by Testi et al. (\citeyear{Testi2014}) revealed that the typical $a_{max}$ value may also differ from clump to clump (see Figure \ref{fig:beta_hist3bins}). This result suggests that dust grains can grow significantly as a clump itself evolves.}

\end{enumerate}

Following Sadavoy et al. (\citeyear{Sadavoy2013}), we mapped \textbeta \space values over several star-forming clump at angular resolutions of $\sim0.5$ arcminutes. Our study is the first of its kind to extend such analysis to the major star-forming clumps across a molecular cloud. In Perseus, we found \textbeta \ values which are significantly different from the diffuse ISM values (i.e., $\sim2$), indicative of dust grain evolution in the denser, star-forming ISM. The discovery of coherent small-scale, low-\textbeta \space regions, and their coincidence with local temperature peaks have provided some hints on the origins of these low-\textbeta \space value grains. Further high-resolution observations of \textbeta \space toward protostellar cores and starless cores would be most welcome. Observations of potential sublimation of ice mantles may also help to constrain the role of ice mantles. 

\acknowledgments
This work was possible with funding from the Natural Sciences and Engineering Research Council of Canada (NSERC) Postgraduate Scholarships. We acknowledge the support
by NSERC via Discovery grants, and the National Research Council of Canada (NRC). We wish to recognize and acknowledge the very significant cultural role and reverence that the summit of Maunakea has always had within the indigenous Hawaiian community. We are most fortunate to have the opportunity to conduct observations from this mountain. 

The James Clerk Maxwell Telescope has historically been operated by the Joint Astronomy Centre on behalf of the Science and Technology Facilities Council of the United Kingdom, the NRC, and the Netherlands Organization for Scientific Research. Additional funds for the construction of SCUBA-2 were provided by the Canada Foundation for Innovation. We thank the JCMT staff for their support of the GBS team in data collection and reduction efforts. 

This research has made use of NASA's Astrophysics Data System (ADS) and the facilities of the Canadian Astronomy Data Centre (CADC) operated by the NRC with the support of the Canadian Space Agency (CSA). We have made use of the SIMBAD database (\citealt{Wenger2000}), operated at CDS, Strasbourg, France. The \emph{Starlink} software (\citealt{Currie2014}) is supported by the East Asian Observatory. This research made use of \textit{APLpy}, an open-source plotting package for Python hosted at \url{http://aplpy.github.com}.

\vspace{5mm}
\facilities{Herschel (SPIRE and PACS), JCMT (SCUBA-2)}

\software{APLpy \citep{APLpy2012}, Astropy \citep{Astropy2013}, NumPy \citep{NumPy2011}, SciPy \citep{SciPy2001}, Starlink \citep{Currie2014}, Matplotlib \citep{Matplotlib2007}, Python} 

\appendix

\section{Anti-correlated \textbeta \ - $T_d$ Uncertainties}
\label{apdx:betaT_antiCorr}

Anti-correlation between \textbeta \space and $T_d$ can be an artifact of SED fitting using the minimization of $\chi^2$ method due to \textbeta \space - $T_d$ degeneracy. Indeed, Shetty et al. (\citeyear{Shetty2009_II}; \citeyear{Shetty2009_I}) demonstrated that SED fitting using the minimization of $\chi^2$ method can lead to such an anti-correlation when modest amounts of noise are present in data. Such anti-correlations due to erroneous fittings have also been noted in previous works (e.g., \citealt{Keene1980}; \citealt{Blain2003}; \citealt{Sajina2006}; \citealt{Kelly2012}; \citealt{Sadavoy2013}). 

The anti-correlated \textbeta \ - $T_d$ uncertainties arise naturally from both \textbeta \space and $T_d$'s ability to shift the peak of a modified blackbody SED (see Equation \ref{eq:modBBF}). While \textbeta \ is responsible for producing the power-law slope of the Rayleigh-Jeans tail of an SED, it can also shift the SED peak independent of $T_d$. An underestimation of \textbeta, for example, leads to a shallower power-law tail that pulls the SED peak lower in frequency. $T_d$, on the other hand, shifts the SED peak through Wien's displacement law, which acts on the blackbody component of the SED. The position of the SED peak can thus be maintained at a constant value by anti-correlating \textbeta \ and $T_d$, resulting in a degeneracy in reduced $\chi^2$ SED fitting when the height of the SED can be arbitrary scaled by a free $\tau$ parameter. 

\begin{figure}
\centering
\includegraphics[width=0.5\textwidth]{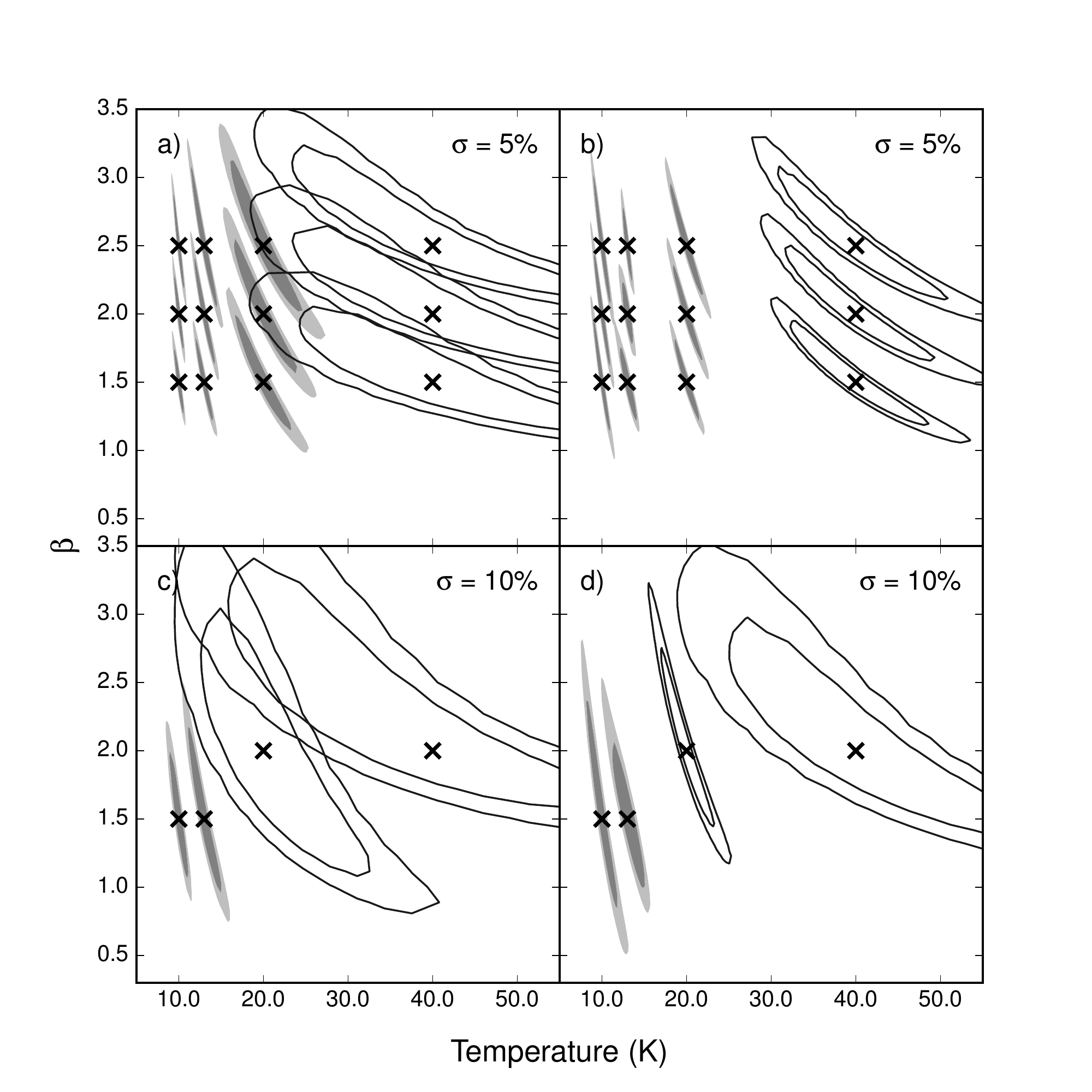}
\caption{The 50\% and 75\% probability contours of $T_d$ and \textbeta \ drawn from SED fits to noisy modified blackbody models using the Monte Carlo approach. The true $T_d$ and \textbeta \ values behind the models are marked with the ``X'' symbols. The modelled SEDs are sampled at 160 \textmu m, 250 \textmu m, 350 \textmu m, 500 \textmu m, and 850 \textmu m for panels a) and c), and at 100 \textmu m, 200 \textmu m, 260 \textmu m, 360 \textmu m, and 580 \textmu m for panels b) and d). The adopted noise levels are shown in each panel.}
  \label{fig:probCont_fromNoise}
\end{figure}

Figure \ref{fig:probCont_fromNoise} shows the 50\% and 75\% probability contours of $T_d$ and \textbeta \ drawn from SED fits to noisy modified blackbody models using the Monte Carlo approach with flux uncertainties of $\sigma = 5\%$ and $10\%$. The true $T_d$ and \textbeta \ values behind the models are marked with the ``X" symbols. The modelled SEDs are sampled at the same wavelengths as our \textit{Herschel} and JCMT bands for the fits shown in panels a) and c), and at the wavelengths used by Shetty et al. (\citeyear{Shetty2009_II}, i.e., 100 \textmu m, 200 \textmu m, 260 \textmu m, 360 \textmu m, and 580 \textmu m) based on observations made by Dupac et al. (\citeyear{Dupac2003}) for the fits shown in panels b) and d). Consistent with the results of Shetty et al. (\citeyear{Shetty2009_II}), anti-correlated uncertainties between $T_d$ and \textbeta \ are significant for SED fits to these wavelengths, even when the flux noise is relatively modest ($\sigma = 5\%$). These uncertainties tend to increase with temperature as the SED peak shifts upwards away from the sampled wavelengths. The fits to SEDs are poorly constrained at $T_d \gtrsim 30$ K for a noise level of $\sigma =5\%$ and at $T_d \gtrsim 20$ K for $\sigma =10\%$.

By having a broader wavelength coverage and a better sampling of the longer wavelengths than those used by Dupac et al. (\citeyear{Dupac2003}), we are able to achieve smaller $T_d$ and \textbeta \ uncertainties at lower temperatures ($T_d \sim 10$ K; see Figure \ref{fig:probCont_fromNoise}) \textbf{at a given noise level}. With 160 \textmu m as our shortest wavelength band instead of the 100 \textmu m used by Dupac et al., however, our uncertainties due to flux noise at higher temperatures are larger than those of Dupac et al. due to having fewer samples on the SED peak. The preferential sampling of the Rayleigh-Jeans tail of the SED instead of the peak, nevertheless, is advantageous in making SED fits more robust against temperature variation along the line of sight (\citealt{Shetty2009_I}).

\begin{figure}
\centering
\includegraphics[width=0.5\textwidth]{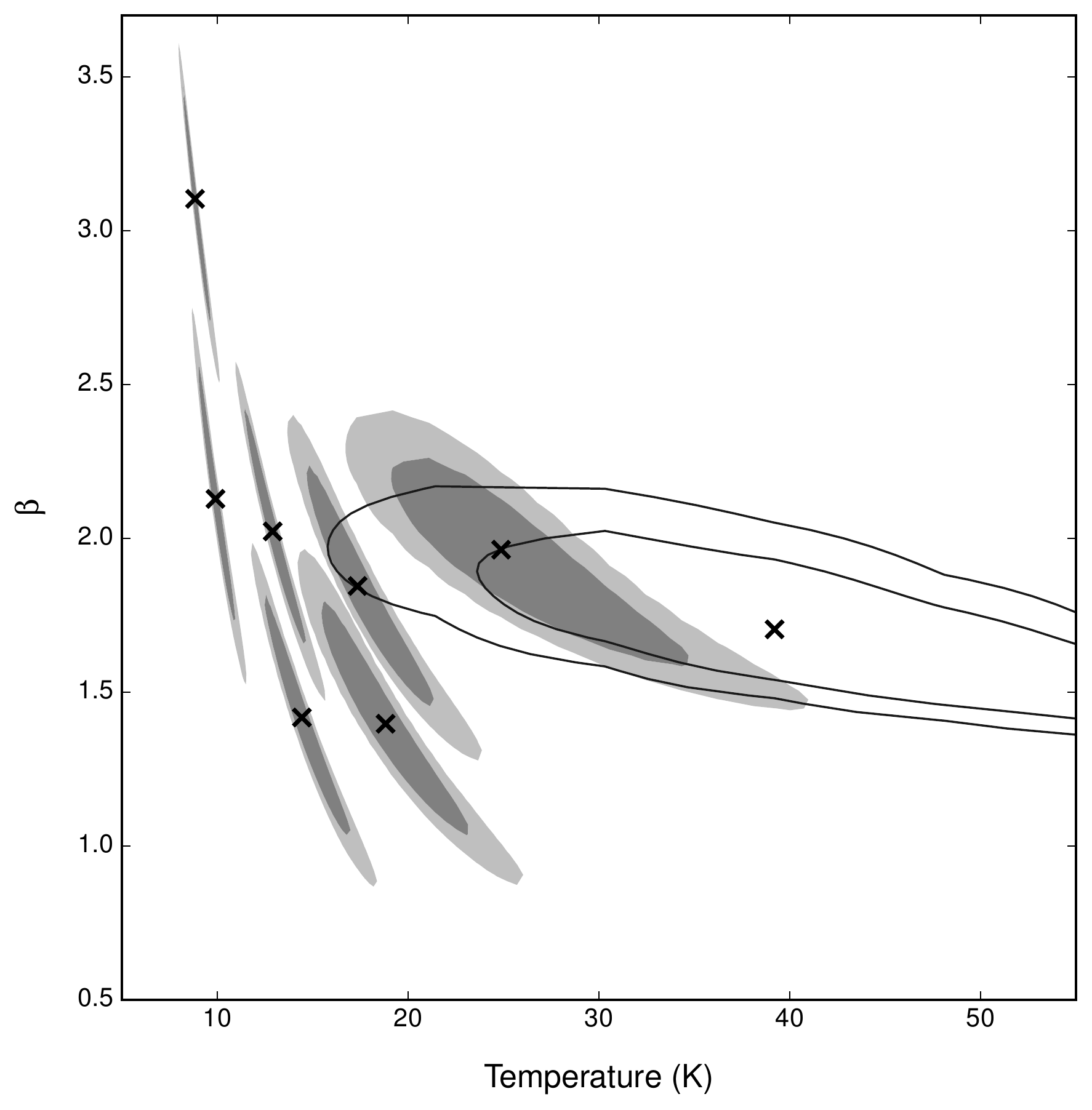}
\caption{The 50\% and 75\% probability contours of $T_d$ and \textbeta \ drawn from SED fits to a few selected pixels in our NGC 1333 data using the Monte Carlo approach. The uncertainties in our data are assumed to be the dominated by flux calibration errors and random flux calibration corrections are generated for each instrument (i.e., PACS, SPIRE, and SCUBA-2) with an adopted error of 10\%. The best fit $T_d$ and \textbeta \ values are marked with ``X'' symbols.}
  \label{fig:probCont_betaTemp_ngc1333}
\end{figure}

The median flux noises in our data over where SED fits are made are small ($< 1\%$ in \textit{Herschel} bands and $\sim 5\%$ in SCUBA-2 850 \textmu m band) relative to our assumed flux calibration uncertainties \textbf{($20\%$ for PACS 160 \textmu m band and $10\%$ for the other bands)}. \textbf{Indeed, we found the inclusion of measured rms noise in addition to the flux calibration uncertainties (correlated within each instrument) to have a negligible effect on the $T_d$ and \textbeta \ uncertainties derived using the Monte Carlo method with the NGC 1333 data.} We therefore assumed flux noise to be negligible \textbf{with respect to the flux calibration uncertainties, which are systematic across a map, in our final uncertainties analysis.} Due the color calibration uncertainties being of the same order as the flux noise, we also assumed the color calibration error to be negligible. 

Figure \ref{fig:probCont_betaTemp_ngc1333} shows the 50\% and 75\% probability contours of $T_d$ and \textbeta \ drawn from SED fits to a few selected pixels in our NGC 1333 data using the Monte Carlo approach. \textbf{Instead of adding random noise to each \textbf{individual} band, random flux calibration corrections are generated for each instrument and the bands within each instrument received the same flux correction at each iteration of the Monte Carlo simulation. We adopted a flux calibration error of 20\% for PACS, 10\% for SPIRE, and 10\% for SCUBA-2.} The best fit $T_d$ and \textbeta \ values are marked with ``X'' symbols. The $T_d$ and \textbeta \ uncertainties due to flux calibration we quoted in this paper are effectively derived from fitting a Gaussian to these probability contours collapsed onto one of these axises (see \citealt{Sadavoy2013} for details).

As shown in Figure \ref{fig:probCont_betaTemp_ngc1333}, even though the probability contours of these selected pixels collectively resemble the shape of the $T_d$ - \textbeta \ scatter plot of NGC 1333 shown in Figure \ref{fig:beta_vs_temp_Per}d, the probability contours of each pixel are relatively small in comparison to the overall distribution. \textbf{Furthermore, due to the flux calibration uncertainties being systematic across a map, erroneous offsets in our derived parameters should be spatially correlated on a scale of $\sim 30'$, which is the size of our smallest maps (i.e., SCUBA-2 maps). The relative pixel to pixel uncertainties between the derived parameters in each map should therefore be smaller than the stated absolute uncertainties demonstrated in Figure \ref{fig:probCont_betaTemp_ngc1333}.} 

The derived $T_d$ values are well constrained at low temperatures ($\lesssim 20$ K), which include most of our pixels (see Figure \ref{fig:histo_temp_Per}). \textbf{At these low temperatures, the uncertainties in the derived \textbeta \ are relatively large with respective to the range of derived \textbeta \ values, and decrease only slightly at higher temperatures.} Despite these large uncertainties, low, intermediate, and high \textbeta \ values can still be distinguished from each other at a \textbf{1-$\sigma$} confidence interval.

\textbf{Our adopted flux calibration uncertainties for extended sources are relatively conservative compared to the values estimated for point sources ($<  7\%$ for PACS bands, \citealt{Balog2014}; $\sim 5\%$ for SPIRE bands, \citealt{Bendo2013}; $< 8\%$ for SCUBA-2 850 \textmu m band). The most common $T_d$ values found in all Perseus clumps, except for IC 348, agree with each other and with the ammonia gas temperatures of Perseus dense cores (\citealt{Rosolowsky2008}) to within 0.5 K. In comparison, our estimated absolute $T_d$ uncertainties at these temperatures is $\sim 0.8$ K. This result suggests that the flux calibration uncertainties we assumed are indeed relatively conservative.}

\begin{figure}
\centering
\includegraphics[width=0.5\textwidth]{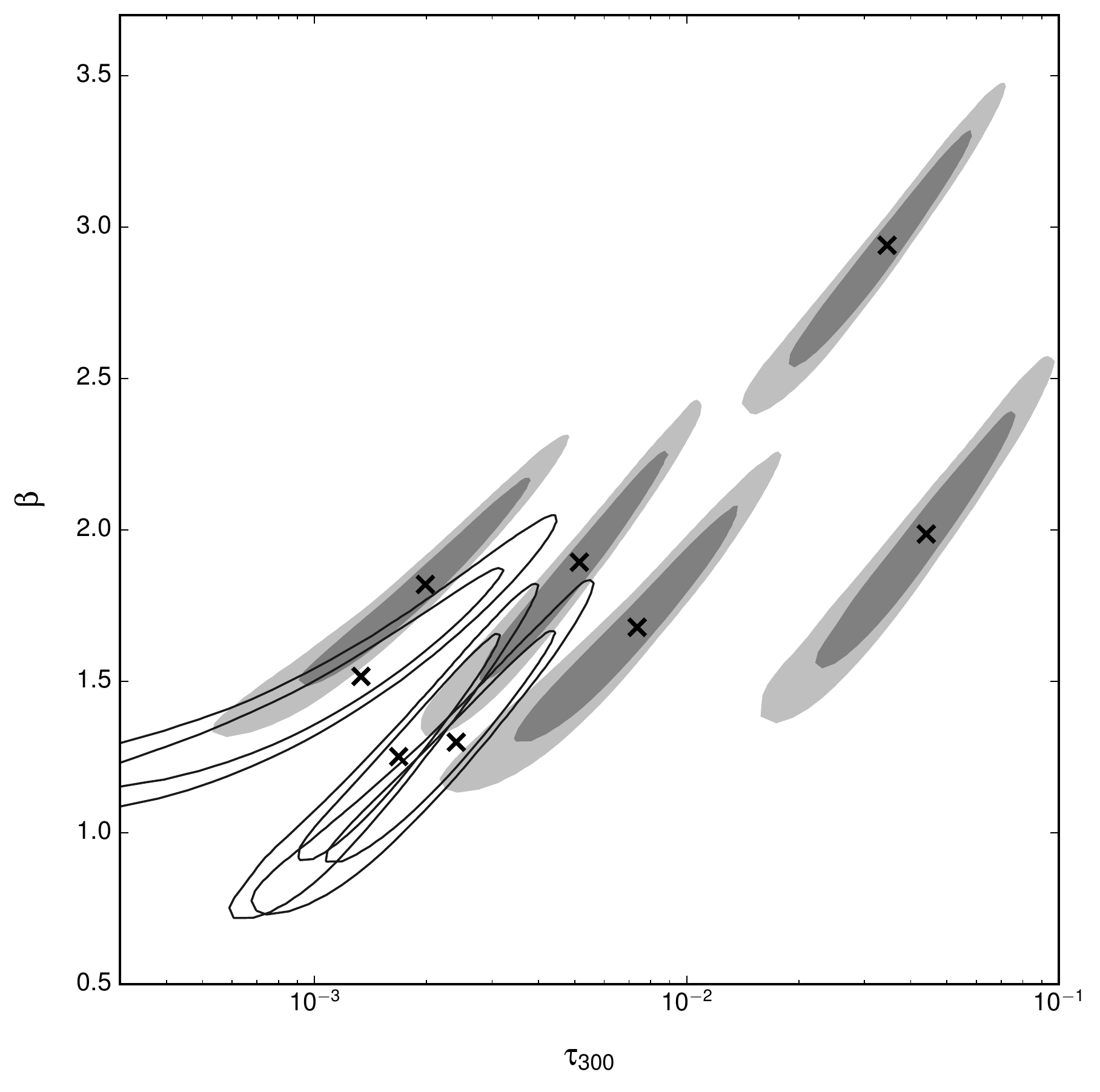}
\caption{The 50\% and 75\% probability contours of $\tau_{300}$ and \textbeta \ drawn from SED fits to the same pixels in NGC 1333 as those in Figure \ref{fig:probCont_betaTemp_ngc1333}, using the same method. The best fit $\tau_{300}$ and \textbeta \ values are marked with ``X'' symbols.}
  \label{fig:probCont_betaTau_ngc1333}
\end{figure}

Figure \ref{fig:probCont_betaTau_ngc1333} shows the 50\% and 75\% probability contours of $\tau_{300}$ and \textbeta \ drawn from SED fits to the same pixels as in Figure \ref{fig:probCont_betaTemp_ngc1333}. As expected, $\tau_{300}$ and \textbeta \ are positively correlated, given that the uncertainties between $T_d$ and \textbeta \ and between $T_d$ and $\tau$ are anti-correlated. The anti-correlation found in the latter is due to the flux of a modified blackbody being positively correlated with $T_d$ (see in Equation \ref{eq:modBBF}). We find the slopes of the correlated uncertainties to be relatively well defined in the log($\tau$) - \textbeta \ space for well constrained SED fits (i.e., at $T_d \lesssim 30$ K), and relatively constant regardless of the fitted parameters. Given these correlated uncertainties, we did not find a significant underlying correlation between \textbeta \ and $\tau_{300}$ shown in Figure \ref{fig:beta_vs_tau300_Per}.
\\
\bibliography{bibliography}

\begin{thebibliography}{}
\expandafter\ifx\csname natexlab\endcsname\relax\def\natexlab#1{#1}\fi

\bibitem[{{Aannestad}(1975)}]{Aannestad1975}
{Aannestad}, P.~A. 1975, \apj, 200, 30

\bibitem[{{Agladze} {et~al.}(1996){Agladze}, {Sievers}, {Jones}, {Burlitch}, \&
  {Beckwith}}]{Agladze1996}
{Agladze}, N.~I., {Sievers}, A.~J., {Jones}, S.~A., {Burlitch}, J.~M., \&
  {Beckwith}, S.~V.~W. 1996, \apj, 462, 1026

\bibitem[{{Andr{\'e}} \& {Saraceno}(2005)}]{Andre2005}
{Andr{\'e}}, P., \& {Saraceno}, P. 2005, in ESA Special Publication, Vol. 577,
  ESA Special Publication, ed. A.~{Wilson}, 179--184

\bibitem[{{Andr{\'e}} {et~al.}(2000){Andr{\'e}}, {Ward-Thompson}, \&
  {Barsony}}]{Andre2000}
{Andr{\'e}}, P., {Ward-Thompson}, D., \& {Barsony}, M. 2000, Protostars and
  Planets IV, 59

\bibitem[{{Andr{\'e}} {et~al.}(2010){Andr{\'e}}, {Men'shchikov}, {Bontemps},
  {K{\"o}nyves}, {Motte}, {Schneider}, {Didelon}, {Minier}, {Saraceno},
  {Ward-Thompson}, {di Francesco}, {White}, {Molinari}, {Testi}, {Abergel},
  {Griffin}, {Henning}, {Royer}, {Mer{\'{\i}}n}, {Vavrek}, {Attard},
  {Arzoumanian}, {Wilson}, {Ade}, {Aussel}, {Baluteau}, {Benedettini},
  {Bernard}, {Blommaert}, {Cambr{\'e}sy}, {Cox}, {di Giorgio}, {Hargrave},
  {Hennemann}, {Huang}, {Kirk}, {Krause}, {Launhardt}, {Leeks}, {Le Pennec},
  {Li}, {Martin}, {Maury}, {Olofsson}, {Omont}, {Peretto}, {Pezzuto}, {Prusti},
  {Roussel}, {Russeil}, {Sauvage}, {Sibthorpe}, {Sicilia-Aguilar}, {Spinoglio},
  {Waelkens}, {Woodcraft}, \& {Zavagno}}]{Andre2010}
{Andr{\'e}}, P., {Men'shchikov}, A., {Bontemps}, S., {et~al.} 2010, \aap, 518,
  L102

\bibitem[{{Anglada}(1996)}]{Anglada1996}
{Anglada}, G. 1996, in Astronomical Society of the Pacific Conference Series,
  Vol.~93, Radio Emission from the Stars and the Sun, ed. A.~R. {Taylor} \&
  J.~M. {Paredes}, 3--14

\bibitem[{{Arab} {et~al.}(2012){Arab}, {Abergel}, {Habart}, {Bernard-Salas},
  {Ayasso}, {Dassas}, {Martin}, \& {White}}]{Arab2012}
{Arab}, H., {Abergel}, A., {Habart}, E., {et~al.} 2012, \aap, 541, A19

\bibitem[{{Arce} {et~al.}(2010){Arce}, {Borkin}, {Goodman}, {Pineda}, \&
  {Halle}}]{Arce2010}
{Arce}, H.~G., {Borkin}, M.~A., {Goodman}, A.~A., {Pineda}, J.~E., \& {Halle},
  M.~W. 2010, \apj, 715, 1170

\bibitem[{{Aspin}(2003)}]{Aspin2003}
{Aspin}, C. 2003, \aj, 125, 1480

\bibitem[{{Astropy Collaboration} {et~al.}(2013){Astropy Collaboration},
  {Robitaille}, {Tollerud}, {Greenfield}, {Droettboom}, {Bray}, {Aldcroft},
  {Davis}, {Ginsburg}, {Price-Whelan}, {Kerzendorf}, {Conley}, {Crighton},
  {Barbary}, {Muna}, {Ferguson}, {Grollier}, {Parikh}, {Nair}, {Unther},
  {Deil}, {Woillez}, {Conseil}, {Kramer}, {Turner}, {Singer}, {Fox}, {Weaver},
  {Zabalza}, {Edwards}, {Azalee Bostroem}, {Burke}, {Casey}, {Crawford},
  {Dencheva}, {Ely}, {Jenness}, {Labrie}, {Lim}, {Pierfederici}, {Pontzen},
  {Ptak}, {Refsdal}, {Servillat}, \& {Streicher}}]{Astropy2013}
{Astropy Collaboration}, {Robitaille}, T.~P., {Tollerud}, E.~J., {et~al.} 2013,
  \aap, 558, A33

\bibitem[{{Bachiller} {et~al.}(1998){Bachiller}, {Guilloteau}, {Gueth},
  {Tafalla}, {Dutrey}, {Codella}, \& {Castets}}]{Bachiller1998}
{Bachiller}, R., {Guilloteau}, S., {Gueth}, F., {et~al.} 1998, \aap, 339, L49

\bibitem[{{Bachiller} {et~al.}(1990){Bachiller}, {Martin-Pintado}, {Tafalla},
  {Cernicharo}, \& {Lazareff}}]{Bachiller1990a}
{Bachiller}, R., {Martin-Pintado}, J., {Tafalla}, M., {Cernicharo}, J., \&
  {Lazareff}, B. 1990, \aap, 231, 174

\bibitem[{{Bally} {et~al.}(1997){Bally}, {Devine}, {Alten}, \&
  {Sutherland}}]{Bally1997}
{Bally}, J., {Devine}, D., {Alten}, V., \& {Sutherland}, R.~S. 1997, \apj, 478,
  603

\bibitem[{{Bally} {et~al.}(2008){Bally}, {Walawender}, {Johnstone}, {Kirk}, \&
  {Goodman}}]{Bally2008}
{Bally}, J., {Walawender}, J., {Johnstone}, D., {Kirk}, H., \& {Goodman}, A.
  2008, {The Perseus Cloud}, ed. B.~{Reipurth}, 308

\bibitem[{{Balog} {et~al.}(2014){Balog}, {M{\"u}ller}, {Nielbock}, {Altieri},
  {Klaas}, {Blommaert}, {Linz}, {Lutz}, {Mo{\'o}r}, {Billot}, {Sauvage}, \&
  {Okumura}}]{Balog2014}
{Balog}, Z., {M{\"u}ller}, T., {Nielbock}, M., {et~al.} 2014, Experimental
  Astronomy, 37, 129

\bibitem[{{Beckwith} \& {Sargent}(1991)}]{Beckwith1991}
{Beckwith}, S.~V.~W., \& {Sargent}, A.~I. 1991, \apj, 381, 250

\bibitem[{{Bendo} {et~al.}(2013){Bendo}, {Griffin}, {Bock}, {Conversi},
  {Dowell}, {Lim}, {Lu}, {North}, {Papageorgiou}, {Pearson}, {Pohlen},
  {Polehampton}, {Schulz}, {Shupe}, {Sibthorpe}, {Spencer}, {Swinyard},
  {Valtchanov}, \& {Xu}}]{Bendo2013}
{Bendo}, G.~J., {Griffin}, M.~J., {Bock}, J.~J., {et~al.} 2013, \mnras, 433,
  3062

\bibitem[{{Bintley} {et~al.}(2014){Bintley}, {Holland}, {MacIntosh}, {Friberg},
  {Bell}, {Berke}, {Berry}, {Berthold}, {Cookson}, {Coulson}, {Currie},
  {Dempsey}, {Gibb}, {Gorges}, {Graves}, {Jenness}, {Johnstone}, {Parsons},
  {Thomas}, {Walther}, \& {Wouterloot}}]{Bintley2014}
{Bintley}, D., {Holland}, W.~S., {MacIntosh}, M.~J., {et~al.} 2014, in
  \procspie, Vol. 9153, Millimeter, Submillimeter, and Far-Infrared Detectors
  and Instrumentation for Astronomy VII, 915303

\bibitem[{{Blain} {et~al.}(2003){Blain}, {Barnard}, \& {Chapman}}]{Blain2003}
{Blain}, A.~W., {Barnard}, V.~E., \& {Chapman}, S.~C. 2003, \mnras, 338, 733

\bibitem[{{Boudet} {et~al.}(2005){Boudet}, {Mutschke}, {Nayral}, {J{\"a}ger},
  {Bernard}, {Henning}, \& {Meny}}]{Boudet2005}
{Boudet}, N., {Mutschke}, H., {Nayral}, C., {et~al.} 2005, \apj, 633, 272

\bibitem[{{Buckle} {et~al.}(2009){Buckle}, {Hills}, {Smith}, {Dent}, {Bell},
  {Curtis}, {Dace}, {Gibson}, {Graves}, {Leech}, {Richer}, {Williamson},
  {Withington}, {Yassin}, {Bennett}, {Hastings}, {Laidlaw}, {Lightfoot},
  {Burgess}, {Dewdney}, {Hovey}, {Willis}, {Redman}, {Wooff}, {Berry},
  {Cavanagh}, {Davis}, {Dempsey}, {Friberg}, {Jenness}, {Kackley}, {Rees},
  {Tilanus}, {Walther}, {Zwart}, {Klapwijk}, {Kroug}, \&
  {Zijlstra}}]{Buckle2009}
{Buckle}, J.~V., {Hills}, R.~E., {Smith}, H., {et~al.} 2009, \mnras, 399, 1026

\bibitem[{{Cernis}(1990)}]{Cernis1990}
{Cernis}, K. 1990, \apss, 166, 315

\bibitem[{{Chandler} {et~al.}(1998){Chandler}, {Barsony}, \&
  {Moore}}]{Chandler1998}
{Chandler}, C.~J., {Barsony}, M., \& {Moore}, T.~J.~T. 1998, \mnras, 299, 789

\bibitem[{{Chandler} \& {Richer}(2000)}]{Chandler2000}
{Chandler}, C.~J., \& {Richer}, J.~S. 2000, \apj, 530, 851

\bibitem[{{Chapin} {et~al.}(2013){Chapin}, {Berry}, {Gibb}, {Jenness}, {Scott},
  {Tilanus}, {Economou}, \& {Holland}}]{Chapin2013}
{Chapin}, E.~L., {Berry}, D.~S., {Gibb}, A.~G., {et~al.} 2013, \mnras, 430,
  2545

\bibitem[{{Chen}(2015)}]{Chen2015MsT}
{Chen}, M.~C. 2015, Master's thesis, University of Victoria

\bibitem[{{Chiang} {et~al.}(2012){Chiang}, {Looney}, \& {Tobin}}]{Chiang2012}
{Chiang}, H.-F., {Looney}, L.~W., \& {Tobin}, J.~J. 2012, \apj, 756, 168

\bibitem[{{Choi} {et~al.}(2011){Choi}, {Kang}, {Tatematsu}, {Lee}, \&
  {Park}}]{Choi2011}
{Choi}, M., {Kang}, M., {Tatematsu}, K., {Lee}, J.-E., \& {Park}, G. 2011,
  \pasj, 63, 1281

\bibitem[{{Connelley} {et~al.}(2008){Connelley}, {Reipurth}, \&
  {Tokunaga}}]{Connelley2008}
{Connelley}, M.~S., {Reipurth}, B., \& {Tokunaga}, A.~T. 2008, \aj, 135, 2496

\bibitem[{{Currie} {et~al.}(2014){Currie}, {Berry}, {Jenness}, {Gibb}, {Bell},
  \& {Draper}}]{Currie2014}
{Currie}, M.~J., {Berry}, D.~S., {Jenness}, T., {et~al.} 2014, in Astronomical
  Society of the Pacific Conference Series, Vol. 485, Astronomical Data
  Analysis Software and Systems XXIII, ed. N.~{Manset} \& P.~{Forshay}, 391

\bibitem[{{Drabek} {et~al.}(2012){Drabek}, {Hatchell}, {Friberg}, {Richer},
  {Graves}, {Buckle}, {Nutter}, {Johnstone}, \& {Di Francesco}}]{Drabek2012}
{Drabek}, E., {Hatchell}, J., {Friberg}, P., {et~al.} 2012, \mnras, 426, 23

\bibitem[{{Draine} \& {Lee}(1984)}]{Draine1984}
{Draine}, B.~T., \& {Lee}, H.~M. 1984, \apj, 285, 89

\bibitem[{{Draine} {et~al.}(1983){Draine}, {Roberge}, \&
  {Dalgarno}}]{Draine1983}
{Draine}, B.~T., {Roberge}, W.~G., \& {Dalgarno}, A. 1983, \apj, 264, 485

\bibitem[{{Dunham} {et~al.}(2015){Dunham}, {Allen}, {Evans},
  {Broekhoven-Fiene}, {Cieza}, {Di Francesco}, {Gutermuth}, {Harvey},
  {Hatchell}, {Heiderman}, {Huard}, {Johnstone}, {Kirk}, {Matthews}, {Miller},
  {Peterson}, \& {Young}}]{Dunham2015}
{Dunham}, M.~M., {Allen}, L.~E., {Evans}, II, N.~J., {et~al.} 2015, \apjs, 220,
  11

\bibitem[{{Dupac} {et~al.}(2003){Dupac}, {Bernard}, {Boudet}, {Giard},
  {Lamarre}, {M{\'e}ny}, {Pajot}, {Ristorcelli}, {Serra}, {Stepnik}, \&
  {Torre}}]{Dupac2003}
{Dupac}, X., {Bernard}, J.-P., {Boudet}, N., {et~al.} 2003, \aap, 404, L11

\bibitem[{{Enoch} {et~al.}(2009){Enoch}, {Evans}, {Sargent}, \&
  {Glenn}}]{Enoch2009}
{Enoch}, M.~L., {Evans}, II, N.~J., {Sargent}, A.~I., \& {Glenn}, J. 2009,
  \apj, 692, 973

\bibitem[{{Evans} {et~al.}(2001){Evans}, {Rawlings}, {Shirley}, \&
  {Mundy}}]{Evans2001}
{Evans}, II, N.~J., {Rawlings}, J.~M.~C., {Shirley}, Y.~L., \& {Mundy}, L.~G.
  2001, \apj, 557, 193

\bibitem[{{Friesen} {et~al.}(2005){Friesen}, {Johnstone}, {Naylor}, \&
  {Davis}}]{Friesen2005}
{Friesen}, R.~K., {Johnstone}, D., {Naylor}, D.~A., \& {Davis}, G.~R. 2005,
  \mnras, 361, 460

\bibitem[{{Griffin} {et~al.}(2010){Griffin}, {Abergel}, {Abreu}, {Ade},
  {Andr{\'e}}, {Augueres}, {Babbedge}, {Bae}, {Baillie}, {Baluteau}, {Barlow},
  {Bendo}, {Benielli}, {Bock}, {Bonhomme}, {Brisbin}, {Brockley-Blatt},
  {Caldwell}, {Cara}, {Castro-Rodriguez}, {Cerulli}, {Chanial}, {Chen},
  {Clark}, {Clements}, {Clerc}, {Coker}, {Communal}, {Conversi}, {Cox},
  {Crumb}, {Cunningham}, {Daly}, {Davis}, {de Antoni}, {Delderfield}, {Devin},
  {di Giorgio}, {Didschuns}, {Dohlen}, {Donati}, {Dowell}, {Dowell}, {Duband},
  {Dumaye}, {Emery}, {Ferlet}, {Ferrand}, {Fontignie}, {Fox}, {Franceschini},
  {Frerking}, {Fulton}, {Garcia}, {Gastaud}, {Gear}, {Glenn}, {Goizel},
  {Griffin}, {Grundy}, {Guest}, {Guillemet}, {Hargrave}, {Harwit}, {Hastings},
  {Hatziminaoglou}, {Herman}, {Hinde}, {Hristov}, {Huang}, {Imhof}, {Isaak},
  {Israelsson}, {Ivison}, {Jennings}, {Kiernan}, {King}, {Lange}, {Latter},
  {Laurent}, {Laurent}, {Leeks}, {Lellouch}, {Levenson}, {Li}, {Li},
  {Lilienthal}, {Lim}, {Liu}, {Lu}, {Madden}, {Mainetti}, {Marliani}, {McKay},
  {Mercier}, {Molinari}, {Morris}, {Moseley}, {Mulder}, {Mur}, {Naylor},
  {Nguyen}, {O'Halloran}, {Oliver}, {Olofsson}, {Olofsson}, {Orfei}, {Page},
  {Pain}, {Panuzzo}, {Papageorgiou}, {Parks}, {Parr-Burman}, {Pearce},
  {Pearson}, {P{\'e}rez-Fournon}, {Pinsard}, {Pisano}, {Podosek}, {Pohlen},
  {Polehampton}, {Pouliquen}, {Rigopoulou}, {Rizzo}, {Roseboom}, {Roussel},
  {Rowan-Robinson}, {Rownd}, {Saraceno}, {Sauvage}, {Savage}, {Savini},
  {Sawyer}, {Scharmberg}, {Schmitt}, {Schneider}, {Schulz}, {Schwartz},
  {Shafer}, {Shupe}, {Sibthorpe}, {Sidher}, {Smith}, {Smith}, {Smith},
  {Spencer}, {Stobie}, {Sudiwala}, {Sukhatme}, {Surace}, {Stevens}, {Swinyard},
  {Trichas}, {Tourette}, {Triou}, {Tseng}, {Tucker}, {Turner}, {Vaccari},
  {Valtchanov}, {Vigroux}, {Virique}, {Voellmer}, {Walker}, {Ward}, {Waskett},
  {Weilert}, {Wesson}, {White}, {Whitehouse}, {Wilson}, {Winter}, {Woodcraft},
  {Wright}, {Xu}, {Zavagno}, {Zemcov}, {Zhang}, \& {Zonca}}]{Griffin2010}
{Griffin}, M.~J., {Abergel}, A., {Abreu}, A., {et~al.} 2010, \aap, 518, L3

\bibitem[{{Hatchell} {et~al.}(2005){Hatchell}, {Richer}, {Fuller},
  {Qualtrough}, {Ladd}, \& {Chandler}}]{Hatchell2005}
{Hatchell}, J., {Richer}, J.~S., {Fuller}, G.~A., {et~al.} 2005, \aap, 440, 151

\bibitem[{{Hatchell} {et~al.}(2013){Hatchell}, {Wilson}, {Drabek}, {Curtis},
  {Richer}, {Nutter}, {Di Francesco}, {Ward-Thompson}, \& {JCMT GBS
  Consortium}}]{Hatchell2013}
{Hatchell}, J., {Wilson}, T., {Drabek}, E., {et~al.} 2013, \mnras, 429, L10

\bibitem[{{Henning} {et~al.}(1995){Henning}, {Michel}, \&
  {Stognienko}}]{Henning1995}
{Henning}, T., {Michel}, B., \& {Stognienko}, R. 1995, \planss, 43, 1333

\bibitem[{{Herbig}(1998)}]{Herbig1998}
{Herbig}, G.~H. 1998, \apj, 497, 736

\bibitem[{{Herbst}(2008)}]{Herbst2008}
{Herbst}, W. 2008, {Star Formation in IC 348}, ed. B.~{Reipurth}, 372

\bibitem[{{Hildebrand}(1983)}]{Hildebrand1983}
{Hildebrand}, R.~H. 1983, \qjras, 24, 267

\bibitem[{{Hirota} {et~al.}(2008){Hirota}, {Bushimata}, {Choi}, {Honma},
  {Imai}, {Iwadate}, {Jike}, {Kameya}, {Kamohara}, {Kan-Ya}, {Kawaguchi},
  {Kijima}, {Kobayashi}, {Kuji}, {Kurayama}, {Manabe}, {Miyaji}, {Nagayama},
  {Nakagawa}, {Oh}, {Omodaka}, {Oyama}, {Sakai}, {Sasao}, {Sato}, {Shibata},
  {Tamura}, \& {Yamashita}}]{Hirota2008}
{Hirota}, T., {Bushimata}, T., {Choi}, Y.~K., {et~al.} 2008, \pasj, 60, 37

\bibitem[{{Holland} {et~al.}(2013){Holland}, {Bintley}, {Chapin},
  {Chrysostomou}, {Davis}, {Dempsey}, {Duncan}, {Fich}, {Friberg}, {Halpern},
  {Irwin}, {Jenness}, {Kelly}, {MacIntosh}, {Robson}, {Scott}, {Ade},
  {Atad-Ettedgui}, {Berry}, {Craig}, {Gao}, {Gibb}, {Hilton}, {Hollister},
  {Kycia}, {Lunney}, {McGregor}, {Montgomery}, {Parkes}, {Tilanus}, {Ullom},
  {Walther}, {Walton}, {Woodcraft}, {Amiri}, {Atkinson}, {Burger}, {Chuter},
  {Coulson}, {Doriese}, {Dunare}, {Economou}, {Niemack}, {Parsons},
  {Reintsema}, {Sibthorpe}, {Smail}, {Sudiwala}, \& {Thomas}}]{Holland2013}
{Holland}, W.~S., {Bintley}, D., {Chapin}, E.~L., {et~al.} 2013, \mnras, 430,
  2513

\bibitem[{{Hollenbach} \& {McKee}(1989)}]{Hollenbach1989}
{Hollenbach}, D., \& {McKee}, C.~F. 1989, \apj, 342, 306

\bibitem[{Hunter(2007)}]{Matplotlib2007}
Hunter, J.~D. 2007, Computing in Science Engineering, 9, 90

\bibitem[{{Ivezic} {et~al.}(1999){Ivezic}, {Nenkova}, \&
  {Elitzur}}]{Ivezic1999}
{Ivezic}, Z., {Nenkova}, M., \& {Elitzur}, M. 1999, ArXiv Astrophysics
  e-prints, astro-ph/9910475

\bibitem[{{Jenness} {et~al.}(2011){Jenness}, {Berry}, {Chapin}, {Economou},
  {Gibb}, \& {Scott}}]{Jenness2011}
{Jenness}, T., {Berry}, D., {Chapin}, E., {et~al.} 2011, in Astronomical
  Society of the Pacific Conference Series, Vol. 442, Astronomical Data
  Analysis Software and Systems XX, ed. I.~N. {Evans}, A.~{Accomazzi}, D.~J.
  {Mink}, \& A.~H. {Rots}, 281

\bibitem[{{Jennings} {et~al.}(1987){Jennings}, {Cameron}, {Cudlip}, \&
  {Hirst}}]{Jennings1987}
{Jennings}, R.~E., {Cameron}, D.~H.~M., {Cudlip}, W., \& {Hirst}, C.~J. 1987,
  \mnras, 226, 461

\bibitem[{Jones {et~al.}(2001--)Jones, Oliphant, Peterson,
  {et~al.}}]{SciPy2001}
Jones, E., Oliphant, T., Peterson, P., {et~al.} 2001--, {SciPy}: Open source
  scientific tools for {Python}, [Online; accessed 2016-04-04]

\bibitem[{{J{\o}rgensen} {et~al.}(2006){J{\o}rgensen}, {Johnstone}, {van
  Dishoeck}, \& {Doty}}]{Jorgensen2006b}
{J{\o}rgensen}, J.~K., {Johnstone}, D., {van Dishoeck}, E.~F., \& {Doty}, S.~D.
  2006, \aap, 449, 609

\bibitem[{Kackley {et~al.}(2010)Kackley, Scott, Chapin, \&
  Friberg}]{Kackley2010}
Kackley, R., Scott, D., Chapin, E., \& Friberg, P. 2010, JCMT Telescope Control
  System upgrades for SCUBA-2, doi:10.1117/12.857397

\bibitem[{{Keene} {et~al.}(1980){Keene}, {Hildebrand}, {Whitcomb}, \&
  {Harper}}]{Keene1980}
{Keene}, J., {Hildebrand}, R.~H., {Whitcomb}, S.~E., \& {Harper}, D.~A. 1980,
  \apjl, 240, L43

\bibitem[{{Kelly} {et~al.}(2012){Kelly}, {Shetty}, {Stutz}, {Kauffmann},
  {Goodman}, \& {Launhardt}}]{Kelly2012}
{Kelly}, B.~C., {Shetty}, R., {Stutz}, A.~M., {et~al.} 2012, \apj, 752, 55

\bibitem[{{K{\"o}nyves} {et~al.}(2015){K{\"o}nyves}, {Andr{\'e}},
  {Men'shchikov}, {Palmeirim}, {Arzoumanian}, {Schneider}, {Roy}, {Didelon},
  {Maury}, {Shimajiri}, {Di Francesco}, {Bontemps}, {Peretto}, {Benedettini},
  {Bernard}, {Elia}, {Griffin}, {Hill}, {Kirk}, {Ladjelate}, {Marsh}, {Martin},
  {Motte}, {Nguy{\^e}n Luong}, {Pezzuto}, {Roussel}, {Rygl}, {Sadavoy},
  {Schisano}, {Spinoglio}, {Ward-Thompson}, \& {White}}]{Konyves2015}
{K{\"o}nyves}, V., {Andr{\'e}}, P., {Men'shchikov}, A., {et~al.} 2015, \aap,
  584, A91

\bibitem[{{Kwon} {et~al.}(2009){Kwon}, {Looney}, {Mundy}, {Chiang}, \&
  {Kemball}}]{Kwon2009}
{Kwon}, W., {Looney}, L.~W., {Mundy}, L.~G., {Chiang}, H.-F., \& {Kemball},
  A.~J. 2009, \apj, 696, 841

\bibitem[{{Lefloch} {et~al.}(1998){Lefloch}, {Castets}, {Cernicharo}, {Langer},
  \& {Zylka}}]{Lefloch1998}
{Lefloch}, B., {Castets}, A., {Cernicharo}, J., {Langer}, W.~D., \& {Zylka}, R.
  1998, \aap, 334, 269

\bibitem[{{Martin} {et~al.}(2012){Martin}, {Roy}, {Bontemps},
  {Miville-Desch{\^e}nes}, {Ade}, {Bock}, {Chapin}, {Devlin}, {Dicker},
  {Griffin}, {Gundersen}, {Halpern}, {Hargrave}, {Hughes}, {Klein}, {Marsden},
  {Mauskopf}, {Netterfield}, {Olmi}, {Patanchon}, {Rex}, {Scott}, {Semisch},
  {Truch}, {Tucker}, {Tucker}, {Viero}, \& {Wiebe}}]{Martin2012}
{Martin}, P.~G., {Roy}, A., {Bontemps}, S., {et~al.} 2012, \apj, 751, 28

\bibitem[{{Mennella} {et~al.}(1998){Mennella}, {Brucato}, {Colangeli},
  {Palumbo}, {Rotundi}, \& {Bussoletti}}]{Mennella1998}
{Mennella}, V., {Brucato}, J.~R., {Colangeli}, L., {et~al.} 1998, \apj, 496,
  1058

\bibitem[{{Miotello} {et~al.}(2014){Miotello}, {Testi}, {Lodato}, {Ricci},
  {Rosotti}, {Brooks}, {Maury}, \& {Natta}}]{Miotello2014}
{Miotello}, A., {Testi}, L., {Lodato}, G., {et~al.} 2014, \aap, 567, A32

\bibitem[{{Miyake} \& {Nakagawa}(1993)}]{Miyake1993}
{Miyake}, K., \& {Nakagawa}, Y. 1993, Icarus, 106, 20

\bibitem[{{Moriarty-Schieven} {et~al.}(2006){Moriarty-Schieven}, {Johnstone},
  {Bally}, \& {Jenness}}]{Moriarty-Schieven2006}
{Moriarty-Schieven}, G.~H., {Johnstone}, D., {Bally}, J., \& {Jenness}, T.
  2006, \apj, 645, 357

\bibitem[{{Ossenkopf} \& {Henning}(1994)}]{Ossenkopf1994}
{Ossenkopf}, V., \& {Henning}, T. 1994, \aap, 291, 943

\bibitem[{{Ott}(2010)}]{Ott2010}
{Ott}, S. 2010, in Astronomical Society of the Pacific Conference Series, Vol.
  434, Astronomical Data Analysis Software and Systems XIX, ed. Y.~{Mizumoto},
  K.-I. {Morita}, \& M.~{Ohishi}, 139

\bibitem[{{Pezzuto} {et~al.}(2012){Pezzuto}, {Elia}, {Schisano}, {Strafella},
  {Di Francesco}, {Sadavoy}, {Andr{\'e}}, {Benedettini}, {Bernard}, {di
  Giorgio}, {Facchini}, {Hennemann}, {Hill}, {K{\"o}nyves}, {Molinari},
  {Motte}, {Nguyen-Luong}, {Peretto}, {Pestalozzi}, {Polychroni}, {Rygl},
  {Saraceno}, {Schneider}, {Spinoglio}, {Testi}, {Ward-Thompson}, \&
  {White}}]{Pezzuto2012}
{Pezzuto}, S., {Elia}, D., {Schisano}, E., {et~al.} 2012, \aap, 547, A54

\bibitem[{{Planck Collaboration} {et~al.}(2011){Planck Collaboration},
  {Abergel}, {Ade}, {Aghanim}, {Arnaud}, {Ashdown}, {Aumont}, {Baccigalupi},
  {Balbi}, {Banday}, {Barreiro}, {Bartlett}, {Battaner}, {Benabed},
  {Beno{\^i}t}, {Bernard}, {Bersanelli}, {Bhatia}, {Bock}, {Bonaldi}, {Bond},
  {Borrill}, {Bouchet}, {Boulanger}, {Bucher}, {Burigana}, {Cabella},
  {Cardoso}, {Catalano}, {Cay{\'o}n}, {Challinor}, {Chamballu}, {Chiang},
  {Chiang}, {Christensen}, {Clements}, {Colombi}, {Couchot}, {Coulais},
  {Crill}, {Cuttaia}, {Danese}, {Davies}, {Davis}, {de Bernardis}, {de
  Gasperis}, {de Rosa}, {de Zotti}, {Delabrouille}, {Delouis}, {D{\'e}sert},
  {Dickinson}, {Dobashi}, {Donzelli}, {Dor{\'e}}, {D{\"o}rl}, {Douspis},
  {Dupac}, {Efstathiou}, {En{\ss}lin}, {Eriksen}, {Finelli}, {Forni},
  {Frailis}, {Franceschi}, {Galeotta}, {Ganga}, {Giard}, {Giardino},
  {Giraud-H{\'e}raud}, {Gonz{\'a}lez-Nuevo}, {G{\'o}rski}, {Gratton},
  {Gregorio}, {Gruppuso}, {Guillet}, {Hansen}, {Harrison},
  {Henrot-Versill{\'e}}, {Herranz}, {Hildebrandt}, {Hivon}, {Hobson}, {Holmes},
  {Hovest}, {Hoyland}, {Huffenberger}, {Jaffe}, {Jones}, {Jones}, {Juvela},
  {Keih{\"a}nen}, {Keskitalo}, {Kisner}, {Kneissl}, {Knox}, {Kurki-Suonio},
  {Lagache}, {Lamarre}, {Lasenby}, {Laureijs}, {Lawrence}, {Leach}, {Leonardi},
  {Leroy}, {Linden-V{\o}rnle}, {L{\'o}pez-Caniego}, {Lubin},
  {Mac{\'{\i}}as-P{\'e}rez}, {MacTavish}, {Maffei}, {Mandolesi}, {Mann},
  {Maris}, {Marshall}, {Martin}, {Mart{\'{\i}}nez-Gonz{\'a}lez}, {Masi},
  {Matarrese}, {Matthai}, {Mazzotta}, {McGehee}, {Meinhold}, {Melchiorri},
  {Mendes}, {Mennella}, {Mitra}, {Miville-Desch{\^e}nes}, {Moneti}, {Montier},
  {Morgante}, {Mortlock}, {Munshi}, {Murphy}, {Naselsky}, {Natoli},
  {Netterfield}, {N{\o}rgaard-Nielsen}, {Noviello}, {Novikov}, {Novikov},
  {Osborne}, {Pajot}, {Paladini}, {Pasian}, {Patanchon}, {Perdereau},
  {Perotto}, {Perrotta}, {Piacentini}, {Piat}, {Plaszczynski}, {Pointecouteau},
  {Polenta}, {Ponthieu}, {Poutanen}, {Pr{\'e}zeau}, {Prunet}, {Puget}, {Reach},
  {Rebolo}, {Reinecke}, {Renault}, {Ricciardi}, {Riller}, {Ristorcelli},
  {Rocha}, {Rosset}, {Rubi{\~n}o-Mart{\'{\i}}n}, {Rusholme}, {Sandri},
  {Santos}, {Savini}, {Scott}, {Seiffert}, {Shellard}, {Smoot}, {Starck},
  {Stivoli}, {Stolyarov}, {Sudiwala}, {Sygnet}, {Tauber}, {Terenzi},
  {Toffolatti}, {Tomasi}, {Torre}, {Tristram}, {Tuovinen}, {Umana},
  {Valenziano}, {Verstraete}, {Vielva}, {Villa}, {Vittorio}, {Wade}, {Wandelt},
  {Yvon}, {Zacchei}, \& {Zonca}}]{PlanckCollaboration2011}
{Planck Collaboration}, {Abergel}, A., {Ade}, P.~A.~R., {et~al.} 2011, \aap,
  536, A25

\bibitem[{{Plunkett} {et~al.}(2013){Plunkett}, {Arce}, {Corder}, {Mardones},
  {Sargent}, \& {Schnee}}]{Plunkett2013}
{Plunkett}, A.~L., {Arce}, H.~G., {Corder}, S.~A., {et~al.} 2013, \apj, 774, 22

\bibitem[{{Poglitsch} {et~al.}(2010){Poglitsch}, {Waelkens}, {Geis},
  {Feuchtgruber}, {Vandenbussche}, {Rodriguez}, {Krause}, {Renotte}, {van
  Hoof}, {Saraceno}, {Cepa}, {Kerschbaum}, {Agn{\`e}se}, {Ali}, {Altieri},
  {Andreani}, {Augueres}, {Balog}, {Barl}, {Bauer}, {Belbachir}, {Benedettini},
  {Billot}, {Boulade}, {Bischof}, {Blommaert}, {Callut}, {Cara}, {Cerulli},
  {Cesarsky}, {Contursi}, {Creten}, {De Meester}, {Doublier}, {Doumayrou},
  {Duband}, {Exter}, {Genzel}, {Gillis}, {Gr{\"o}zinger}, {Henning},
  {Herreros}, {Huygen}, {Inguscio}, {Jakob}, {Jamar}, {Jean}, {de Jong},
  {Katterloher}, {Kiss}, {Klaas}, {Lemke}, {Lutz}, {Madden}, {Marquet},
  {Martignac}, {Mazy}, {Merken}, {Montfort}, {Morbidelli}, {M{\"u}ller},
  {Nielbock}, {Okumura}, {Orfei}, {Ottensamer}, {Pezzuto}, {Popesso},
  {Putzeys}, {Regibo}, {Reveret}, {Royer}, {Sauvage}, {Schreiber}, {Stegmaier},
  {Schmitt}, {Schubert}, {Sturm}, {Thiel}, {Tofani}, {Vavrek}, {Wetzstein},
  {Wieprecht}, \& {Wiezorrek}}]{Poglitsch2010}
{Poglitsch}, A., {Waelkens}, C., {Geis}, N., {et~al.} 2010, \aap, 518, L2

\bibitem[{{Quillen} {et~al.}(2005){Quillen}, {Thorndike}, {Cunningham},
  {Frank}, {Gutermuth}, {Blackman}, {Pipher}, \& {Ridge}}]{Quillen2005}
{Quillen}, A.~C., {Thorndike}, S.~L., {Cunningham}, A., {et~al.} 2005, \apj,
  632, 941

\bibitem[{{Robitaille} \& {Bressert}(2012)}]{APLpy2012}
{Robitaille}, T., \& {Bressert}, E. 2012, {APLpy: Astronomical Plotting Library
  in Python}, Astrophysics Source Code Library, ascl:1208.017

\bibitem[{{Rodr{\'{\i}}guez} {et~al.}(1997){Rodr{\'{\i}}guez}, {Anglada}, \&
  {Curiel}}]{Rodriguez1997}
{Rodr{\'{\i}}guez}, L.~F., {Anglada}, G., \& {Curiel}, S. 1997, \apjl, 480,
  L125

\bibitem[{{Rosolowsky} {et~al.}(2008){Rosolowsky}, {Pineda}, {Foster},
  {Borkin}, {Kauffmann}, {Caselli}, {Myers}, \& {Goodman}}]{Rosolowsky2008}
{Rosolowsky}, E.~W., {Pineda}, J.~E., {Foster}, J.~B., {et~al.} 2008, \apjs,
  175, 509

\bibitem[{{Roussel}(2013)}]{Roussel2013}
{Roussel}, H. 2013, \pasp, 125, 1126

\bibitem[{{Sadavoy}(2013)}]{Sadavoy2013PhDT}
{Sadavoy}, S.~I. 2013, PhD thesis, University of Victoria

\bibitem[{{Sadavoy} {et~al.}(2016){Sadavoy}, {Stutz}, {Schnee}, {Mason}, {Di
  Francesco}, \& {Friesen}}]{Sadavoy2016}
{Sadavoy}, S.~I., {Stutz}, A.~M., {Schnee}, S., {et~al.} 2016, \aap, 588, A30

\bibitem[{{Sadavoy} {et~al.}(2012){Sadavoy}, {di Francesco}, {Andr{\'e}},
  {Pezzuto}, {Bernard}, {Bontemps}, {Bressert}, {Chitsazzadeh}, {Fallscheer},
  {Hennemann}, {Hill}, {Martin}, {Motte}, {Nguyen Luong}, {Peretto}, {Reid},
  {Schneider}, {Testi}, {White}, \& {Wilson}}]{Sadavoy2012}
{Sadavoy}, S.~I., {di Francesco}, J., {Andr{\'e}}, P., {et~al.} 2012, \aap,
  540, A10

\bibitem[{{Sadavoy} {et~al.}(2013){Sadavoy}, {Di Francesco}, {Johnstone},
  {Currie}, {Drabek}, {Hatchell}, {Nutter}, {Andr{\'e}}, {Arzoumanian},
  {Benedettini}, {Bernard}, {Duarte-Cabral}, {Fallscheer}, {Friesen},
  {Greaves}, {Hennemann}, {Hill}, {Jenness}, {K{\"o}nyves}, {Matthews},
  {Mottram}, {Pezzuto}, {Roy}, {Rygl}, {Schneider-Bontemps}, {Spinoglio},
  {Testi}, {Tothill}, {Ward-Thompson}, {White}, {JCMT}, \& {Herschel Gould Belt
  Survey Teams}}]{Sadavoy2013}
{Sadavoy}, S.~I., {Di Francesco}, J., {Johnstone}, D., {et~al.} 2013, \apj,
  767, 126

\bibitem[{{Sadavoy} {et~al.}(2014){Sadavoy}, {Di Francesco}, {Andr{\'e}},
  {Pezzuto}, {Bernard}, {Maury}, {Men'shchikov}, {Motte}, {Nguy{\^e}n-Lu'o'ng},
  {Schneider}, {Arzoumanian}, {Benedettini}, {Bontemps}, {Elia}, {Hennemann},
  {Hill}, {K{\"o}nyves}, {Louvet}, {Peretto}, {Roy}, \& {White}}]{Sadavoy2014}
{Sadavoy}, S.~I., {Di Francesco}, J., {Andr{\'e}}, P., {et~al.} 2014, \apjl,
  787, L18

\bibitem[{{Sajina} {et~al.}(2006){Sajina}, {Scott}, {Dennefeld}, {Dole},
  {Lacy}, \& {Lagache}}]{Sajina2006}
{Sajina}, A., {Scott}, D., {Dennefeld}, M., {et~al.} 2006, \mnras, 369, 939

\bibitem[{{Salji} {et~al.}(2015){Salji}, {Richer}, {Buckle}, {Hatchell},
  {Kirk}, {Beaulieu}, {Berry}, {Broekhoven-Fiene}, {Currie}, {Fich}, {Jenness},
  {Johnstone}, {Mottram}, {Nutter}, {Pattle}, {Pineda}, {Quinn}, {Tisi},
  {Walker-Smith}, {Francesco}, {Hogerheijde}, {Ward-Thompson}, {Bastien},
  {Butner}, {Chen}, {Chrysostomou}, {Coude}, {Davis}, {Drabek-Maunder},
  {Duarte-Cabral}, {Fiege}, {Friberg}, {Friesen}, {Fuller}, {Graves},
  {Greaves}, {Gregson}, {Holland}, {Joncas}, {Kirk}, {Knee}, {Mairs}, {Marsh},
  {Matthews}, {Moriarty-Schieven}, {Rawlings}, {Robertson}, {Rosolowsky},
  {Rumble}, {Sadavoy}, {Thomas}, {Tothill}, {Viti}, {White}, {Wilson},
  {Wouterloot}, {Yates}, \& {Zhu}}]{Salji2015a}
{Salji}, C.~J., {Richer}, J.~S., {Buckle}, J.~V., {et~al.} 2015, \mnras, 449,
  1769

\bibitem[{{Sandell} \& {Knee}(2001)}]{Sandell2001}
{Sandell}, G., \& {Knee}, L.~B.~G. 2001, \apjl, 546, L49

\bibitem[{{Schnee} {et~al.}(2014){Schnee}, {Mason}, {Di Francesco}, {Friesen},
  {Li}, {Sadavoy}, \& {Stanke}}]{Schnee2014}
{Schnee}, S., {Mason}, B., {Di Francesco}, J., {et~al.} 2014, \mnras, 444, 2303

\bibitem[{{Schnee} {et~al.}(2009){Schnee}, {Rosolowsky}, {Foster}, {Enoch}, \&
  {Sargent}}]{Schnee2009}
{Schnee}, S., {Rosolowsky}, E., {Foster}, J., {Enoch}, M., \& {Sargent}, A.
  2009, \apj, 691, 1754

\bibitem[{{Schnee} {et~al.}(2010){Schnee}, {Enoch}, {Noriega-Crespo}, {Sayers},
  {Terebey}, {Caselli}, {Foster}, {Goodman}, {Kauffmann}, {Padgett}, {Rebull},
  {Sargent}, \& {Shetty}}]{Schnee2010}
{Schnee}, S., {Enoch}, M., {Noriega-Crespo}, A., {et~al.} 2010, \apj, 708, 127

\bibitem[{{Scoville} \& {Kwan}(1976)}]{Scoville1976}
{Scoville}, N.~Z., \& {Kwan}, J. 1976, \apj, 206, 718

\bibitem[{{Shetty} {et~al.}(2009{\natexlab{a}}){Shetty}, {Kauffmann}, {Schnee},
  \& {Goodman}}]{Shetty2009_II}
{Shetty}, R., {Kauffmann}, J., {Schnee}, S., \& {Goodman}, A.~A.
  2009{\natexlab{a}}, \apj, 696, 676

\bibitem[{{Shetty} {et~al.}(2009{\natexlab{b}}){Shetty}, {Kauffmann}, {Schnee},
  {Goodman}, \& {Ercolano}}]{Shetty2009_I}
{Shetty}, R., {Kauffmann}, J., {Schnee}, S., {Goodman}, A.~A., \& {Ercolano},
  B. 2009{\natexlab{b}}, \apj, 696, 2234

\bibitem[{{Shirley} {et~al.}(2011){Shirley}, {Mason}, {Mangum}, {Bolin},
  {Devlin}, {Dicker}, \& {Korngut}}]{Shirley2011}
{Shirley}, Y.~L., {Mason}, B.~S., {Mangum}, J.~G., {et~al.} 2011, \aj, 141, 39

\bibitem[{{Shirley} {et~al.}(2005){Shirley}, {Nordhaus}, {Grcevich}, {Evans},
  {Rawlings}, \& {Tatematsu}}]{Shirley2005}
{Shirley}, Y.~L., {Nordhaus}, M.~K., {Grcevich}, J.~M., {et~al.} 2005, \apj,
  632, 982

\bibitem[{{Snell} \& {Edwards}(1981)}]{Snell1981}
{Snell}, R.~L., \& {Edwards}, S. 1981, \apj, 251, 103

\bibitem[{{Swinyard} {et~al.}(2010){Swinyard}, {Ade}, {Baluteau}, {Aussel},
  {Barlow}, {Bendo}, {Benielli}, {Bock}, {Brisbin}, {Conley}, {Conversi},
  {Dowell}, {Dowell}, {Ferlet}, {Fulton}, {Glenn}, {Glauser}, {Griffin},
  {Griffin}, {Guest}, {Imhof}, {Isaak}, {Jones}, {King}, {Leeks}, {Levenson},
  {Lim}, {Lu}, {Makiwa}, {Naylor}, {Nguyen}, {Oliver}, {Panuzzo},
  {Papageorgiou}, {Pearson}, {Pohlen}, {Polehampton}, {Pouliquen},
  {Rigopoulou}, {Ronayette}, {Roussel}, {Rykala}, {Savini}, {Schulz},
  {Schwartz}, {Shupe}, {Sibthorpe}, {Sidher}, {Smith}, {Spencer}, {Trichas},
  {Triou}, {Valtchanov}, {Wesson}, {Woodcraft}, {Xu}, {Zemcov}, \&
  {Zhang}}]{Swinyard2010}
{Swinyard}, B.~M., {Ade}, P., {Baluteau}, J.-P., {et~al.} 2010, \aap, 518, L4

\bibitem[{{Testi} {et~al.}(2014){Testi}, {Birnstiel}, {Ricci}, {Andrews},
  {Blum}, {Carpenter}, {Dominik}, {Isella}, {Natta}, {Williams}, \&
  {Wilner}}]{Testi2014}
{Testi}, L., {Birnstiel}, T., {Ricci}, L., {et~al.} 2014, Protostars and
  Planets VI, 339

\bibitem[{{Tibbs} {et~al.}(2011){Tibbs}, {Flagey}, {Paladini}, {Compi{\`e}gne},
  {Shenoy}, {Carey}, {Noriega-Crespo}, {Dickinson}, {Ali-Ha{\"i}moud},
  {Casassus}, {Cleary}, {Davies}, {Davis}, {Hirata}, \& {Watson}}]{Tibbs2011}
{Tibbs}, C.~T., {Flagey}, N., {Paladini}, R., {et~al.} 2011, \mnras, 418, 1889

\bibitem[{{Tobin} {et~al.}(2016){Tobin}, {Looney}, {Li}, {Chandler}, {Dunham},
  {Segura-Cox}, {Sadavoy}, {Melis}, {Harris}, {Kratter}, \&
  {Perez}}]{Tobin2016}
{Tobin}, J.~J., {Looney}, L.~W., {Li}, Z.-Y., {et~al.} 2016, \apj, 818, 73

\bibitem[{van~der Walt {et~al.}(2011)van~der Walt, Colbert, \&
  Varoquaux}]{NumPy2011}
van~der Walt, S., Colbert, S.~C., \& Varoquaux, G. 2011, Computing in Science
  Engineering, 13, 22

\bibitem[{{Ward-Thompson} {et~al.}(2007){Ward-Thompson}, {Di Francesco},
  {Hatchell}, {Hogerheijde}, {Nutter}, {Bastien}, {Basu}, {Bonnell}, {Bowey},
  {Brunt}, {Buckle}, {Butner}, {Cavanagh}, {Chrysostomou}, {Curtis}, {Davis},
  {Dent}, {van Dishoeck}, {Edmunds}, {Fich}, {Fiege}, {Fissel}, {Friberg},
  {Friesen}, {Frieswijk}, {Fuller}, {Gosling}, {Graves}, {Greaves}, {Helmich},
  {Hills}, {Holland}, {Houde}, {Jayawardhana}, {Johnstone}, {Joncas}, {Kirk},
  {Kirk}, {Knee}, {Matthews}, {Matthews}, {Matzner}, {Moriarty-Schieven},
  {Naylor}, {Padman}, {Plume}, {Rawlings}, {Redman}, {Reid}, {Richer},
  {Shipman}, {Simpson}, {Spaans}, {Stamatellos}, {Tsamis}, {Viti}, {Weferling},
  {White}, {Whitworth}, {Wouterloot}, {Yates}, \& {Zhu}}]{Ward-Thompson2007}
{Ward-Thompson}, D., {Di Francesco}, J., {Hatchell}, J., {et~al.} 2007, \pasp,
  119, 855

\bibitem[{{Wenger} {et~al.}(2000){Wenger}, {Ochsenbein}, {Egret}, {Dubois},
  {Bonnarel}, {Borde}, {Genova}, {Jasniewicz}, {Lalo{\"e}}, {Lesteven}, \&
  {Monier}}]{Wenger2000}
{Wenger}, M., {Ochsenbein}, F., {Egret}, D., {et~al.} 2000, \aaps, 143, 9

\bibitem[{{Wolfire} \& {Churchwell}(1994)}]{Wolfire1994}
{Wolfire}, M.~G., \& {Churchwell}, E. 1994, \apj, 427, 889

\bibitem[{{Yang} \& {Phillips}(2007)}]{Yang2007}
{Yang}, M., \& {Phillips}, T. 2007, \apj, 662, 284

\end{thebibliography}

\end{document}